\documentclass[fleqn,usenatbib]{mnras}

\usepackage{newtxtext,newtxmath}
\usepackage{ulem}

\usepackage[T1]{fontenc}

\DeclareRobustCommand{\VAN}[3]{#2}
\let\VANthebibliography\thebibliography
\def\thebibliography{\DeclareRobustCommand{\VAN}[3]{##3}\VANthebibliography}

\usepackage{graphicx}	
\usepackage{amsmath}	
\usepackage{siunitx}
\usepackage{multirow}
\usepackage{xcolor}
\usepackage[version=4]{mhchem}
\usepackage{soul}	

\DeclareSIUnit{\angstrom}{\textup{\AA}}

\title[Glycolaldehyde Formation by ISM Ice]{Glycolaldehyde Formation Mediated by Interstellar Amorphous Ice: A Computational Study}

\author[M. A. M. Paiva et al.]{
M. A. M. Paiva,$^{1}$
S. Pilling,$^{2}$
E. Mendoza,$^{3}$
B. R. L. Galvão,$^{4}$
H. A. De Abreu$^{1}$\thanks{E-mail: heitorabreu@ufmg.br}
\\
$^{1}$Departamento de Química, ICEx, Universidade Federal de Minas Gerais, Belo Horizonte-MG, 31270-901, Brazil\\
$^{2}$Instituto de Pesquisa e Desenvolvimento (IP\&D), Universidade do Vale do Paraíba (UNIVAP), São José dos Campos 12244-000, São Paulo, Brazil\\
$^{3}$Dpto. Ciencias Integradas y Centro de Estudios Avanzados en Física, Matemáticas y Computación, Universidad de Huelva, 21071 Huelva, Spain\\
$^{4}$Departamento de Química, Centro Federal de Educação Tecnológica de Minas Gerais, 30421-169, Belo Horizonte-MG, Brazil\\
}

\date{Accepted XXX. Received YYY; in original form ZZZ}

\pubyear{2022}

\begin{document}
\label{firstpage}
\pagerange{\pageref{firstpage}--\pageref{lastpage}}
\maketitle

\begin{abstract}
Glycolaldehyde (HOCH$_2$CHO) is the most straightforward sugar detected in the Interstellar Medium (ISM) and participates in the formation pathways of molecules fundamental to life, red such as ribose and derivatives. Although detected in several regions of the ISM, its formation route is still debated and its abundance cannot be explained only by reactions in the gas phase. This work explores a new gas-phase formation mechanism for glycolaldehyde and compares the energy barrier reduction when the same route happens on the surface of amorphous ices. The first step of the mechanism involves the formation of a carbon-carbon bond between formaldehyde (H$_2$CO) and the formyl radical (HCO), with an energy barrier of 27 kJ mol$^{-1}$ (gas-phase). The second step consists of barrierless hydrogen addition. Density functional calculations under periodic boundary conditions were applied to study this reaction path on 10 different amorphous ice surfaces through an Eley-Rideal type mechanism. It was found that the energy barrier is reduced on average by 49\%, leading in some cases to a 100\% reduction. The calculated adsorption energy of glycolaldehyde suggests that it can be promptly desorbed to the gas phase after its formation. This work thus contributes to explaining the detected relative abundances of glycolaldehyde and opens a new methodological framework for studying the formation routes for Complex Organic Molecules (COMs) in interstellar icy grains.

\end{abstract}

\begin{keywords}
astrochemistry -- molecular processes -- ISM: molecules
\end{keywords}



\section{Introduction}

Among the molecules detected in the interstellar medium (ISM), simple and complex organic molecules,  SOM and COMs,\footnote{As discussed in \citet{Herbst2009}, the adjective complex here stands for interstellar molecules containing an organic functional group, as the simple ones, but with more than 5 atoms.} respectively, such as formaldehyde (H$_2$CO), formamide (CH$_3$NO), acetaldehyde (CH$_3$CHO), ethene (H$_2$CCH$_2$), glycolaldehyde (HOCH$_2$CHO), methyl and ethyl formate (HCOOCH$_3$,HCOOCH$_2$CH$_3$), have attracted much attention \citep{herbst2009complex,Lopez2015,zamirri2019quantum}.
In space, COMs have been found in interstellar clouds, hot corinos, hot molecular cores, molecular outflows, circumstellar envelopes around evolved stars, and in low and high mass star forming regions in general \citep{Coletta2020,Nazari2021}. Detections of COMs from young stellar objects and protoplanetary discs suggest the participation of these molecules in the composition of comets and planets in formation, which increase the importance of COMs as the building blocks of the first organic molecules in primitive planets \citep{Walsh2014,fedoseev2015experimental,butscher2015formation}. 

The study of COMs has been an emerging topic in astrochemistry because of their potential role in prebiotic chemistry, {particularly} in the formation of amino acids and other fundamental species for the emergence of life \citep{bulak2021photolysis}. The glycolaldehyde molecule is one of these molecules directly linked to the formation of complex sugars, like ribose and derivatives, essentials to genetic material throughout formose reaction. \citep{woods2013glycolaldehyde,woods2012formation,banfalvi2021prebiotic}. From astronomical observations in the sub-mm domain, \citet{hollis2000interstellar} and \citet{hal2006} carried out a comprehensive study of glycolaldehyde in Sgr~B2 discussing the role of the formose reaction to produce glycolaldehyde from formaldehyde. Recent theoretical and experimental works have investigated again the role of the formose reaction, also known as the Butlerov reaction, under interstellar conditions. \citet{Ahmad2020} studied the role of interstellar H$_2$CO to produce the C$_2$H$_4$O$_2$ isomers: glycolaldehyde, methyl formate and acetic acid (CH$_3$COOH). \citet{Layssac2020} studied experimentally the chemistry of formaldehyde through formose-like reactions. They found that VUV processed interstellar ice analogues containing H$_2$CO can produce sugar-related molecules.

Glycolaldehyde was firstly detected is space in molecular clouds within Sagittarius B2(N) \citep{hollis2000interstellar}, and later on it has been observed in several other different regions, such as the  hot molecular core G31.41+0.31 \citep{beltran2008first}, the Class 0 protostellar binary IRAS 16293-2422 \citep{jorgensen2012detection}, solar-type protostar NGC 1333 IRAS2A \citep{coutens2015detection}, multiple sources of Perseus molecular cloud \citep{de2017glycolaldehyde}, and in shock region L1157-B1 \citep{lefloch2017l1157}. Glycolaldehyde also has been detected in comets like Lovejoy \citep{biver2015ethyl} and 67P/Churyumov–Gerasimenko \citep{goesmann2015organic}. Regarding bioessential sugars,  \citet{Furukawa2019} detected sugar-related compounds in three carbonaceous chondrites demonstrating evidence of extraterrestrial ribose in primitive meteorites. These observations exemplify how widespread the glycolaldehyde molecule is in the ISM environment. 
Despite its importance, its chemical formation routes in space are still not well understood \citep{woods2012formation,woods2013glycolaldehyde,vazart2018ethanol,coutens2018chemical}.

The other two isomers of glycolaldehyde, the methyl formate and acetic acid have also been detected in several astrophysical environments \citep{mehringer1997detection,remijan2003survey,remijan2005survey,shiao2010first}   being all involved in the synthesis of key biomolecules such as glycine (NH$_2$CH$_2$COOH), the simplest amino acid \citep{sorrell2001origin}. Methyl formate is the most abundant of the three isomers detected in space, glycolaldehyde being the less abundant one. Indeed, among the three isomers, methyl formate was the first ever observed outside the Galaxy. \citet{Sewilo2018} reported its first extragalactic detection, along with dimethyl ether in the N113 star forming region in the Large Magellanic Cloud. As discussed by \citet{rachid2017destruction}, such high abundances of methyl formate could be related to its high resistance to ionising radiation field when in solid phase compared with other isomers. However, this species in the gas phase is highly sensitive to ionising radiation \citep{fantuzzi2011photodissociation}. Both authors agreed that besides destruction by radiation, an explanation considering different efficiencies in both production and desorption routes to the gas phase might be invoked in an attempt to explain such observed abundances among isomers. According to \citet{burke2015glycolaldehyde}, the desorption of methyl formate from ice grain mantles occurs at lower temperatures than glycolaldehyde and acetic acid, and at shorter time-scale and therefore the temperature of a region might be an important constrain for the abundances of theses isomers (expected to be mainly produced in ices) to be detected in gas-phase.

In contrast, formaldehyde is a simple molecule found in many regions of space where glycolaldehyde is detected \citep{jorgensen2012detection,leroux2020solid}. By modelling emission signals from the $^{18}$O isotope transitions of the SgrB2(N) region,  an abundance of formaldehyde to glycolaldehyde is estimated in the ratio between 42 and 56 \citep{jorgensen2012detection}. This abundance estimate provides the basis for a possible route of formation of glycolaldehyde from formaldehyde. However, the simple reaction between two formaldehyde molecules does not seem to be favourable even in the solid phase, as shown by \citet{leroux2020solid}, who concluded that the production of glyoxal  followed by fragmentation would occur instead of glycolaldehyde production.

Shortly after the first detection of glycolaldehyde in Sagittarius B2(N), \citet{sorrell2001origin} proposed through theoretical astrophysical models that complex organic molecules such as glycolaldehyde could be formed in the bulk of icy grain mantles with the presence of radicals created by ultraviolet radiation. In this study, it was postulated that reactions between two formyl radical molecules and between formyl radical and methanol could lead to the formation of a glycolaldehyde molecule.

Through theoretical calculations, \citet{jalbout2007can} proposes that gas-phase reactions between two molecules of formaldehyde with subsequent shock with H$_3^+$ have the potential for the final formation of glycolaldehyde and methyl formate. However, the chemical reaction proposed in this work leads to very slow reaction kinetics to produce the abundance of glycolaldehyde detected in the ISM.

\citet{skouteris2018genealogical} proposed, through computational calculations,  reaction pathways for the formation of glycolaldehyde through radical reactions between formaldehyde and hydroxymethyl radical in gas phase. Nevertheless, the reaction is promising with only two elementary steps, it comes up against an energy barrier that makes it clear that glycolaldehyde is not favourable for its formation in the gas phase, unless it develops through ethanol radical precursors.

In a nutshell, astrochemical models indicate that gas-phase routes would be inefficient for the formation of glycolaldehyde, and  grain-surface formation are preferable candidates. Specifically, the mechanism with a formyl radical and methanol molecules with five reaction steps initially proposed by \citet{sorrell2001origin} is the best candidate for matching the observational estimates in low temperatures. However, mechanisms in two steps of two-body reactions involving HCO, H$_2$CO and H look more feasible from a chemistry perspective \citet{sorrell2001origin,beltran2008first,bennett2007formation,woods2013glycolaldehyde}.

 The \citet{bennett2007formation} work shows that a hydrogen atom could add to a CO molecule, generating the formyl radical (HCO) for a posteriori reaction to generate the glycolaldehyde molecules and other isomers. In addition, the study developed by \citet{pantaleone2020chemical} shows that the HCO radical produced in situ by hydrogenation of CO  remains adsorbed at the surface of water ice under ISM conditions. This radical will then be available for further collisions from incoming gas-phase molecules, such as formaldehyde.

Nevertheless, through astrochemical models and theoretical calculations \citet{woods2013glycolaldehyde} showed that the recombination of two formyl radical molecules  is not an efficient initial path for glycolaldehyde formation. In the same study, it was proposed that a surface reaction with an energy barrier of less than 100K would be sufficient to match the observed abundance of glycolaldehyde in the ISM. \citet{chu2016} studied the formation of  glycolaldehyde, ethylene glycol and methyl formate in mixed CO, H$_2$CO and CH$_3$OH ices. They highlighted the importance of radicals as HCO in the formation of COMs. A more recent study using simulations with a gas-grain chemical code for astrochemical modelling came to the same conclusions for recombining two formyl radical molecules \citep{coutens2018chemical}. 

In this work, we perform computational predictions on the possibility of glycolaldehyde formation via reactions between formyl and formaldehyde, which are believed to be important  precursors for this molecule~\citep{beltran2008first,woods2012formation}. This is carried out both in the gas phase and in the presence of amorphous water ice using a new model. 

\section{Computational methods}

To investigate the influence of the ice in catalysing the reaction, the mechanism was initially explored in the gas phase and improved using benchmark calculations. Subsequently, we propose a model for the surface of an amorphous ice, and reevaluate the reaction supported on it in several different ways.

\subsection{Gas Phase}

Geometry optimisations and vibrational calculations  were performed using density functional theory (DFT) with exchange-correlation functional M06-2x and aug-cc-pVQZ (AVQZ) basis set \citep{zhao2008density,peterson2008systematically}. In addition, IRC (Intrinsic Reaction Coordinate) \citep{maeda2015intrinsic,fukui1970formulation,fukui1981path} and NEB (Nudged Elastic Band) \citep{henkelman2000climbing} calculations ensured the correct link between the molecules and the representation of the isomer with the lowest energy. To further improve the accuracy, single point energies were calculated at the M06-2X geometry using the explicitly correlated coupled cluster method with singles and doubles excitation with the third one obtained by perturbative method, with resolution of identity correction (CCSD(T)-F12/RI), using the cc-pVQZ-F12 (VQZ-F12) basis set \citep{knizia2009simplified}. The T1 diagnostic of the various structures were evaluated and are presented in table 1 of the supplementary material. The calculations are performed in the lowest spin state of each molecule. 
We have employed the ORCA program package 5.0.3 version in all gas phase calculations \citep{neese2012orca,neese2018software,neese2020orca}.

\subsection{Solid Phase}

The modelling of an amorphous ice surface is a 
non-trivial task, and there are several possible ways to tackle this problem. One of these is to use clusters of water molecules in the gas phase to mimic the disorganised pattern of a real grain\citep{al2004adsorption,ferrero2020binding}. Alternatively,  crystalline unit cells have also been used as an approximation to the amorphous ones \citep{karssemeijer2014interactions,perrero2021ab,molpeceres2019silicate}. Fewer works have employed amorphous unit cells for periodic calculations, which applied molecular dynamics on crystalline configurations to generate the random pattern \citep{andersson2006molecular,andersson2008photodesorption,oberg2016photochemistry}. 

\begin{figure}
    \centering
	\includegraphics[scale=0.6]{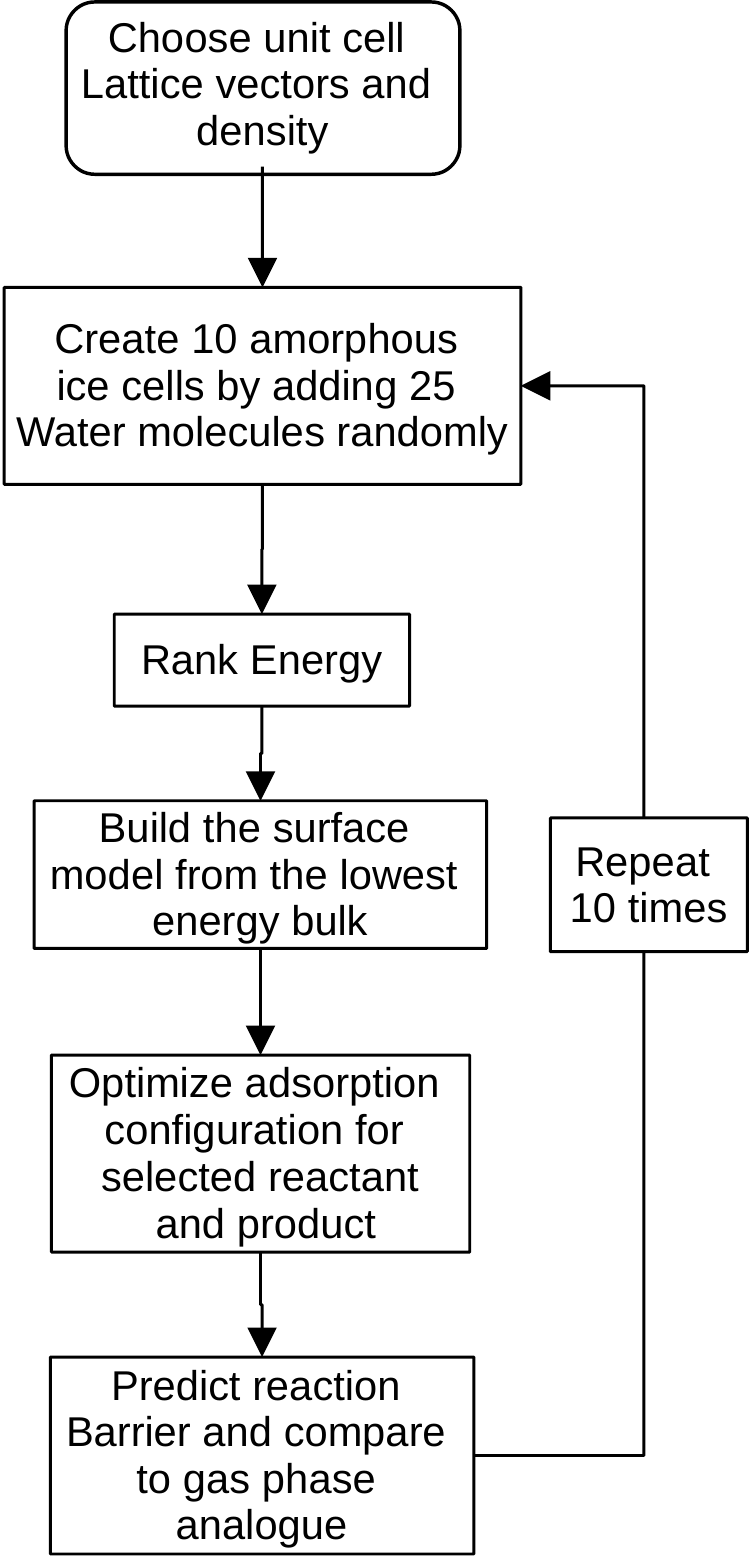}
    \caption{Flow chart of methodology steps used in solid state phase}
    \label{fig:fluxograma1}
\end{figure}

In this work, we propose a faster and yet efficient approach to build representative amorphous unit cells. The main steps of this approach are summarised in Fig.~\ref{fig:fluxograma1}. With a Fortran routine, we create a tetragonal box with dimensions of 10x10x7$\si{\angstrom}$. To match the mean density of ISM amorphous ice, 25 water molecules must be added inside such unit cell.
The addition of each molecule is done by generating three random numbers, which are used to locate the centre of mass of the molecule at a random position inside the cell. Three other random numbers are employed for rotating the molecule around the three Euler angles. This is performed independently for each new molecule, and the only restriction to accept the position of a new one is that its distance to any previously added molecule must be greater than 1.6$\si{\angstrom}$. If this condition is not satisfied, the molecule is discarded and a new random position is tried.
 This process was repeated ten times creating ten different models for the amorphous ice. All obtained structures were optimized, and their total energies were ranked. The cells were classified according to their total energy and only the unit cell with the lowest energy of the group of ten was selected for posterior surface creation.

Surfaces were created by increasing the height of the tetragonal unit cell by 13 $\si{\angstrom}$ resulting in a box of dimensions 10x10x20 $\si{\angstrom}$. Another geometry relaxation was performed at the surface.
As shown in Fig.~\ref{fig:fluxograma1}, this process was repeated ten times, yielding a diverse set of models for the surface of the amorphous ice grain.

All geometry optimizations were performed using DFT calculations with exchange functional PBE using plane waves \citep{perdew1996generalized} with Grimme D3 dispersion correction \citep{grimme2010consistent} and the ultrasoft type pseudopotential \citep{garrity2014pseudopotentials}. The plane wave cutoff energy was set to 40 rydberg (Ry) with a mesh of k points set to 3x3x3 in the bulk and 2x2x1 in the surface models. The reactions involving radicals were of the spin-polarized calculation type, in which unpaired electrons in different orbitals are taken into account.

Subsequently, we proceed to study the adsorption and reaction of the molecules in each of the ten  surfaces generated, thus providing a sample of the many ways the reaction may occur in different ices.  At each stage of the mechanism proposed, reagents, intermediate, and product molecules are positioned on the ten surfaces randomly (following the same routine used for the water molecules in ice, producing random cartesian coordinates near the surface) and structurally optimized. The Eley-Rideal reaction model was used for the mechanism proposed \citep{hama2013surface}, starting with one molecule adsorbed and the other approaching from the gas phase. Two initial routes were tested, one with HCO adsorbed while H$_2$CO approaches from the gas phase and the opposite.

The transition state search is performed using NEB type calculations with 10 images to reproduce the energy barrier expected by the approximation of the HCO and H$_2$CO molecules. Each image is separated by 0.5 \AA\ from the other. To quantify the amount by which the reaction barrier is reduced due to the presence of the ice, a reference calculation employing the same methodology is performed removing the water molecules and with a very large unit cell, thus representing the reaction in gas phase at this calculation level. Note that the benchmark calculations described in the previous section cannot be used as a reference to calculate the barrier height reduction here, as the methods for surface calculation are different. 

It is important to mention that the energy barriers were not correctly described by the PBE functional. An in-depth analysis of this problem can be found in the Results section and more detailed in the supplementary material. It was shown that hybrid functionals are necessary to describe the barrier-height, and to circumvent this difficulty, single-point calculations were performed for each image obtained in the NEB calculations using the B3LYP functional with only the gamma point \citep{stephens1994ab}. Therefore, all energy presented for the calculations including the ice surface are at the B3LYP level with PBE-optimized geometry. We have employed the Quantum Espresso package 7.0 version in all solid phase calculations. \citep{giannozzi2009quantum}.

\section{Results}

In this work, several possible reaction pathways for  glycolaldehyde formation were initially screened, but most were deemed as potentially not important.
The new and efficient route proposed here occurs in two parts: first a formaldehyde molecule reacts with a formyl radical, giving rise to an intermediate product radical, followed by a hydrogenation process leading to the glycolaldehyde formation, such as:

\begin{align}
    \mathrm{H_2CO + HCO\cdot  \rightarrow H_3C_2O_2\cdot} \label{eq1}\\
    \mathrm{H_3C_2O_2\cdot +\  H\cdot \rightarrow H_4C_2O_2} \label{eq2}
\end{align}

Reaction \ref{eq1} is the decisive step and leads to the formation of the carbon-carbon bond. Unlike previous work, here we start with the premise that the HCO and H$_2$CO molecules are stable enough in the ice surface to react with another molecule \citep{pantaleone2020chemical}. First we consider HCO adsorbed in the surface and reacting with formaldehyde coming from the gas phase, and later the opposite situation.

Reaction \ref{eq2} is expected to be fast due to its radical-radical character, and given the abundance of hydrogen atoms in the interstellar medium. The addition of a hydrogen atom promotes the formation of a very stable neutral and closed shell molecule. To understand the catalytic effect of the amorphous ice on this reaction, we first calculate its energy barrier in the gas phase, to later assess if it can be promoted on the surface of a grain.

\subsection{Gas Phase}

\begin{figure}
	\includegraphics[width=\columnwidth]{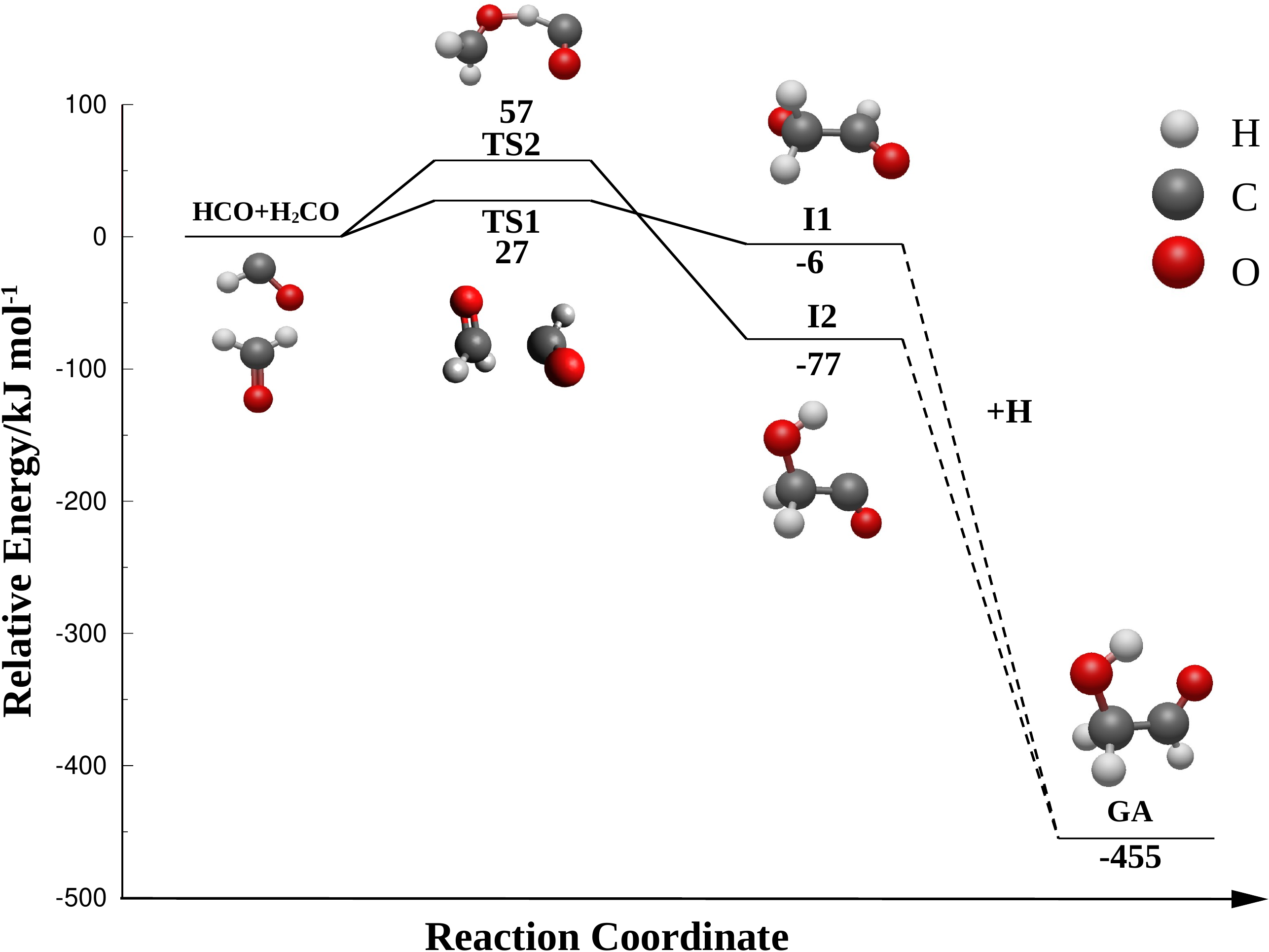}
    \caption{Gas phase mechanisms for  Glycolaldehyde (GA) formation. The solid connections corresponds to reaction (\ref{eq1}) while the dashed line corresponds to reaction (\ref{eq2}). All energies are relative to the hypothetical H+HCO+H$_2$CO separated fragments and were obtained at the CCSD(T)-F12/VQZ-F12 level of theory with ZPE correction.}
    \label{fig:diagram1}
\end{figure}

The results obtained for gas-phase calculations are illustrated in  Fig. \ref{fig:diagram1}. It is shown that the chemical reaction exemplified in equation \ref{eq1} has an activation energy of 27 kJ mol$^{-1}$, probably due to electron repulsion from the molecular orbitals during the intermolecular approximation. As a result of carbon-carbon bond formation, the intermediate $\mathrm{H_3C_2O_2}$ (I1) is produced, which is the glycolaldehyde species without a hydrogen atom, with relative energy of -6 kJ mol$^{-1}$ with respect to the reactants, suggesting that the formation of the carbon-carbon bond slightly stabilises the composite system. The hydrogenation process that occurs in the sequence, as can be seen in Fig. \ref{fig:diagram1}, takes place without an energy barrier resulting in the target molecule glycolaldehyde, with relative energy of -455 kJ mol$^{-1}$. One can note that adding a hydrogen atom leads to a considerable stabilisation of the molecule, as this addition leads to a non-radical diamagnetic system with 12 electrons in a closed shell configuration.

An alternative mechanism of glycolaldehyde formation was also found, with a different direction of approximation between reactant molecules, as can be seen also in Fig. \ref{fig:diagram1}. In this mechanism, the H atom of the HCO radical migrates to the oxygen end of formaldehyde, leading to a transition state (TS2) that is higher in energy than TS1. In contrast, the intermediate formed in this way (I2) has lower energy than I1. The intermediate formation occurs with a hydrogen atom transfer, followed by the carbon-carbon bond formation. In sequence, the formation of the glycolaldehyde molecule occurs via a subsequent step in which the carbon (originally from the HCO molecule) is hydrogenated, and not the oxygen, as in the previous mechanism.

\begin{figure}
	\includegraphics[width=\columnwidth]{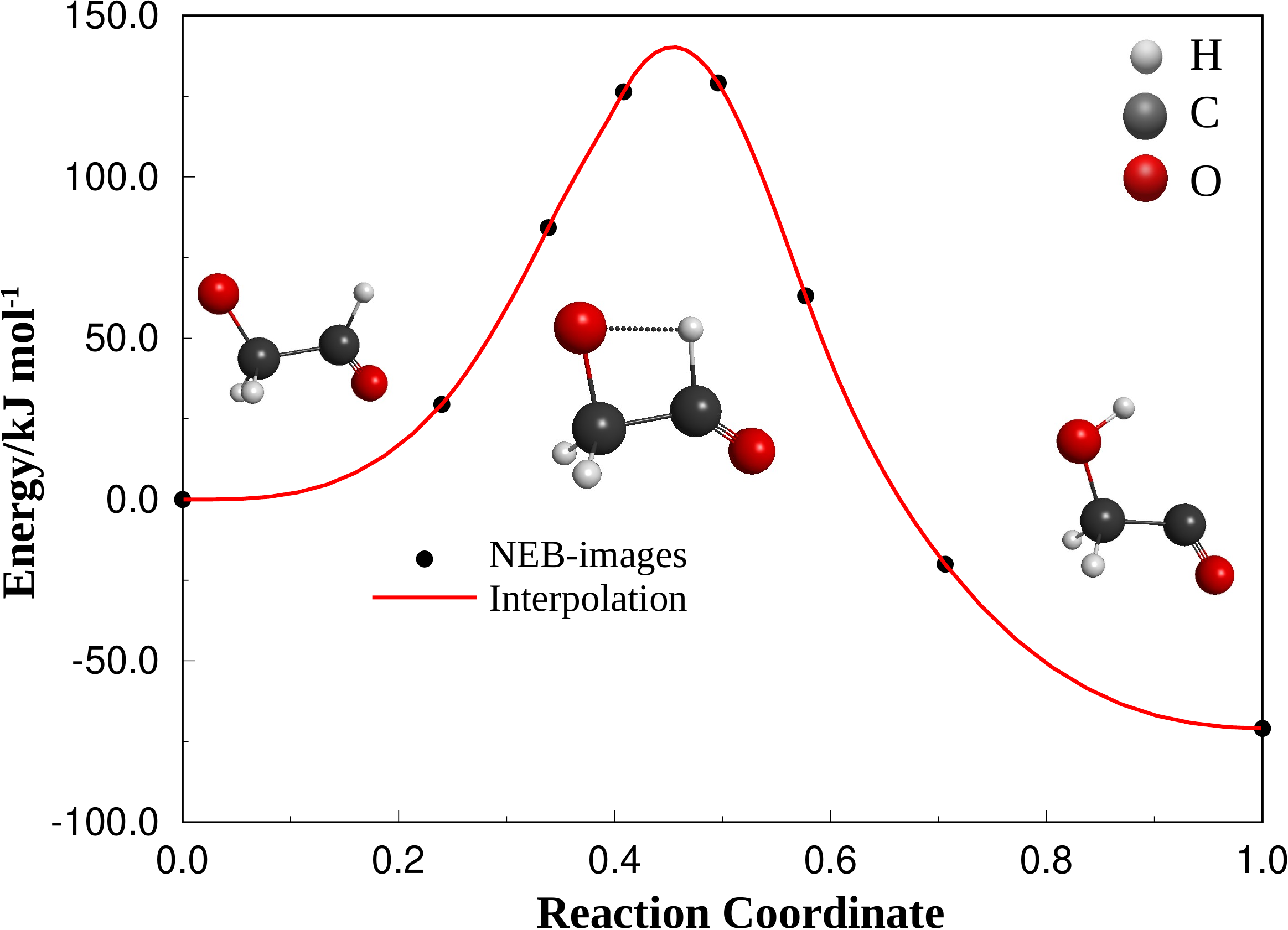}
    \caption{NEB calculation for the possible conversion of intermediate species I1 and I2. Gas phase calculation at M06-2X/aug-cc-pVTZ level.}
    \label{fig:i1i2}
\end{figure}

The interdependence of the mechanisms through TS1 and and TS2 was investigated considering the isomerisation of intermediates I1 and I2, as shown in Fig. \ref{fig:i1i2}. If a possible conversion of the intermediate I1 and I2 occurs, the reaction could go through TS1 to I1 and later isomerize via H-atom migration to I2 before formation of the glycolaldehyde, since I2 has lower energy relative to the reagents. As shown in Fig. \ref{fig:i1i2}, this is not possible because the energy barrier for conversion between the intermediates is too high compared to their  formation barriers. This result indicates that only I1 contributes to glycolaldehyde formation.

Complete energy data at DFT, CCSD(T) and ZPE correction are shown in Table 1 of the supplementary data.

\subsection{Solid Phase}

\subsubsection{Reaction Mechanism}

\begin{figure*}
	\includegraphics[scale=0.3]{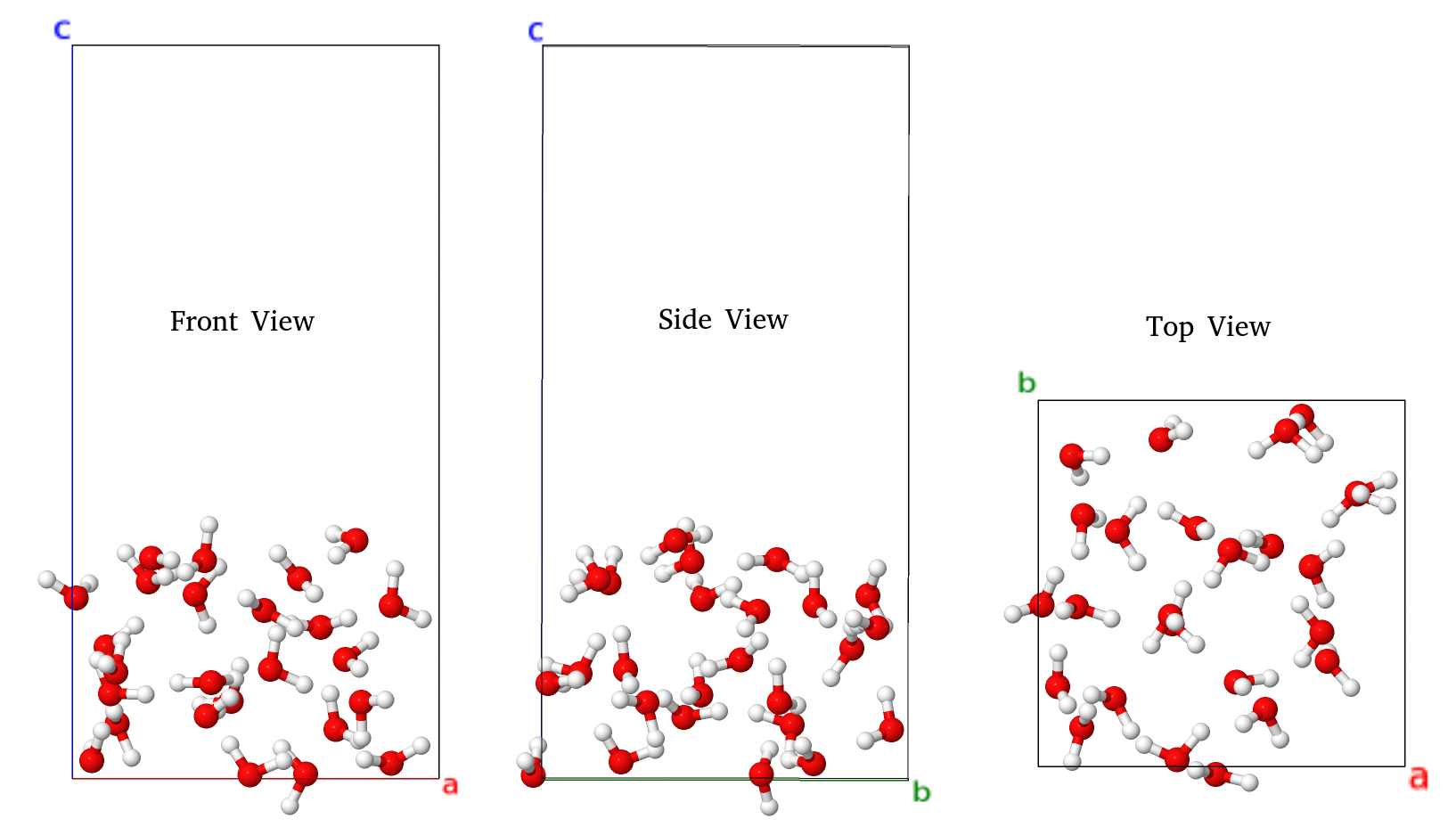}
    \caption{Front, side and top view from one of the surfaces of amorphous ice created in this work}
    \label{fig:super3}
\end{figure*}

Fig. \ref{fig:super3} shows one of the model surfaces of amorphous ice created in this work. The random pattern of water molecules with the presence of non-homogeneous cavities (as expected in an ISM ice) can be observed. The complete group of studied surfaces can be seen in section 3 of the supplementary material. We emphasise here the importance of using a diverse sample of possible surfaces, because the structural arrangement of the site where the reaction takes place may vary substantially in a real amorphous ice grain.

\begin{figure*}
	\includegraphics[scale=0.3]{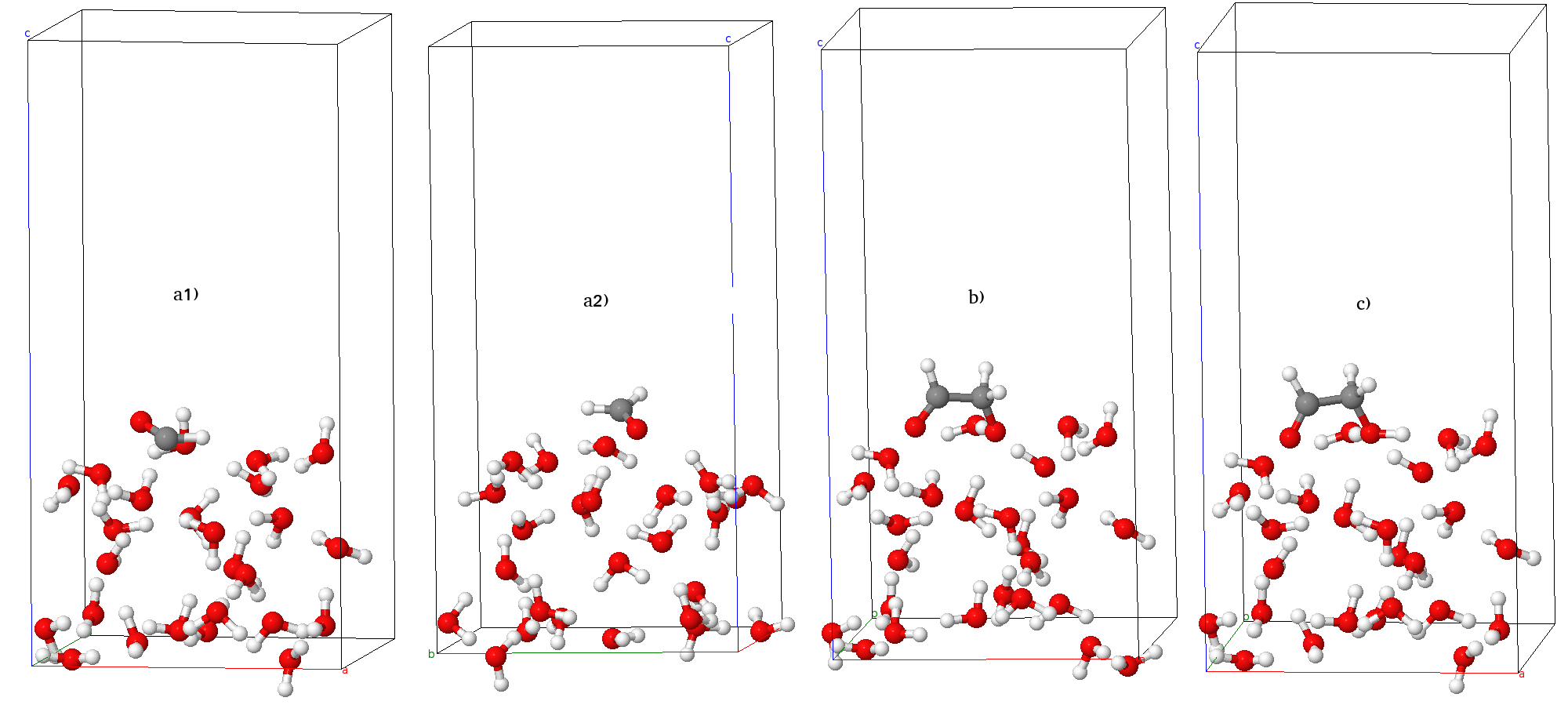}
    \caption{a1) Adsorption of HCO radical a2) Adsorption of H$_2$CO molecule b) Intermediate molecule I1 and c) glycolaldehyde molecule adsorbed in surface of amorphous ice}
    \label{fig:pathexample}
\end{figure*}

The mechanism through TS1 (Fig. \ref{fig:diagram1}) was the most successful in the gas phase, and for this reason it was chosen to be the reference for the surface calculations. We have explored this reaction in two ways: one starting with an H$_2$CO molecule adsorbed at the surface (s-H$_2$CO) with HCO approaching from the gas phase, while the other starts with HCO adsorbed (s-HCO) and being approached by H$_2$CO. Examples of each reactant molecule adsorbed on the surface, as well as the intermediate I1 and the product glycolaldehyde can be seen in Fig. \ref{fig:pathexample}. Since we have 10 possible models for the amorphous ice surface, and the two reaction possibilities described above, this amounts to a total of 20 potential energy barriers (EB) predictions, which are presented in Table \ref{tab:ts1qe}.

To investigate whether glycolaldehyde formation is in fact catalysed when carried out on the surface of amorphous ice, we calculate the decrease in EB provoked by the presence of the surface as $\rm EB^{gas}-EB^{surf}$. It is not possible to carry out benchmark CCSD(T)-F12 with periodic conditions. The reference gas-phase EB was calculated at the same level of theory as the 20 calculations just described, which employ periodic boundary conditions with large unit cells containing only the HCO and H$_2$CO molecule at B3LYP level. 

We should stress here that, to reproduce the benchmark gas-phase EB employing DFT calculations, it was necessary to use hybrid exchange-correlation functionals, since PBE was not able to predict a gas phase barrier. This is true both for localised basis sets and plane wave calculations. The complete exchange-correlation functional study for this system is shown in section 1 of supplementary data.

The results of each different surface studied are shown in Table \ref{tab:ts1qe}. As can be seen, the surface of amorphous ice catalyses the reaction of formation of glycolaldehyde at different levels with an average of 49\% of decrease of the energetic barrier in the s-H$_2$CO path. In some cases, the energetic barrier disappeared such as in surfaces number 4 and 6, leading to a reaction without barrier that should occur very rapidly.

\begin{table}
\centering
\caption{Influence of decrease the energy barrier of the transition state (EB) by ice. Eley-Rideal reaction type starting with H$_2$CO or HCO molecules adsorbed in the surface (s)}
\label{tab:ts1qe}
\begin{tabular}{ccccc}
\hline
\multirow{2}{*}{\#Surface} & \multicolumn{2}{c}{Size of EB (kJ mol$^{-1}$)} & \multicolumn{2}{c}{Decrease of EB (\%)} \\
                           & s-H$_2$CO             & s-HCO                 & s-H$_2$CO          & s-HCO              \\
                           \hline
1                          & 12                    & 0                     & 14                 & 100                \\
2                          & 13                    & 15                    & 5                  & 0                  \\
3                          & 14                    & 30                    & 1                  & 0                  \\
4                          & 0                     & 17                    & 100                & 0                  \\
5                          & 10                    & 18                    & 24                 & 0                  \\
6                          & 0                     & 15                    & 100                & 0                  \\
7                          & 4                     & 14                    & 68                 & 0                  \\
8                          & 15                    & 15                    & 0                  & 0                  \\
9                          & 1                     & 6                     & 91                 & 52                 \\
10                         & 2                     & 27                    & 83                 & 0                 \\
\hline
Average                    & 7                     & 16                    & 49                 & 15                 \\
\hline
\end{tabular}
\end{table}

It can also be seen in Table~\ref{tab:ts1qe} that the path starting from the s-HCO structure is less efficient than that with s-H$_2$CO. In surfaces 1 and 9, there is a complete and partial decrease in the energy barrier. However, on the other surfaces, there is no decrease at all when compared to the gas phase reaction.
It is assumed that this behaviour occurs due to the influence of the surface in stabilising the unpaired electron of the HCO radical, and thus decreasing its reactivity. In the s-H$_2$CO path on the other hand, the adsorption is likely to activate the otherwise stable formaldehyde molecule via an electron density transfer mechanism, and thus catalysing the process of carbon-carbon bond formation.

\begin{figure}
	\includegraphics[width=\columnwidth]{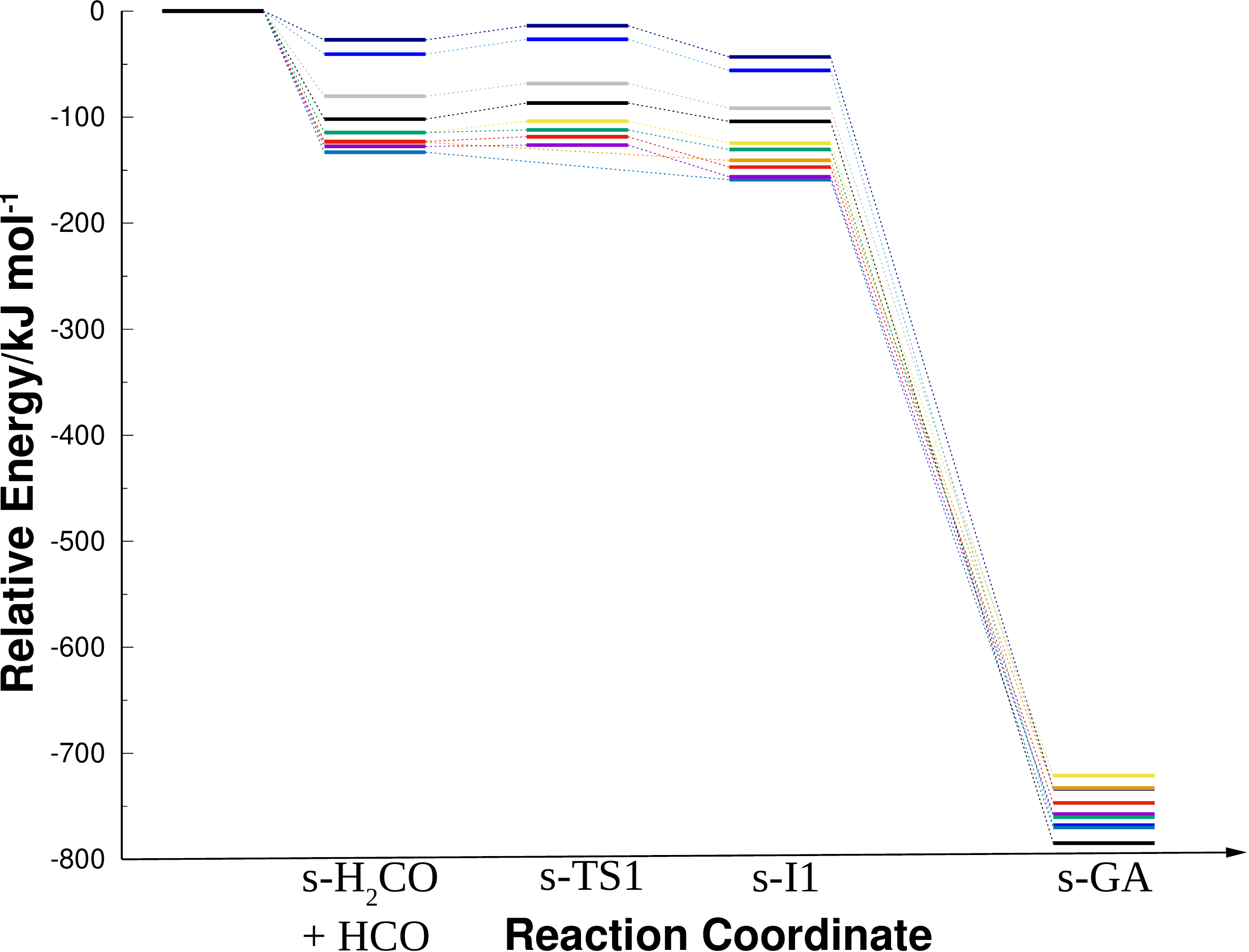}
    \caption{Formation of glycolaldehyde (GA) in the ten model surfaces, each represented by a different colour. The first step corresponds to the adsorption of the H$_2$CO molecule at the surface, followed by HCO attack yielding the intermediate I1 and later the final product. The energy reference in this diagram is the sum of the ice surface energy and the separate reactants. The s- prefix indicates that the molecule is adsorbed on the surface.}
    \label{fig:allsurf1}
\end{figure}

\begin{figure}
	\includegraphics[width=\columnwidth]{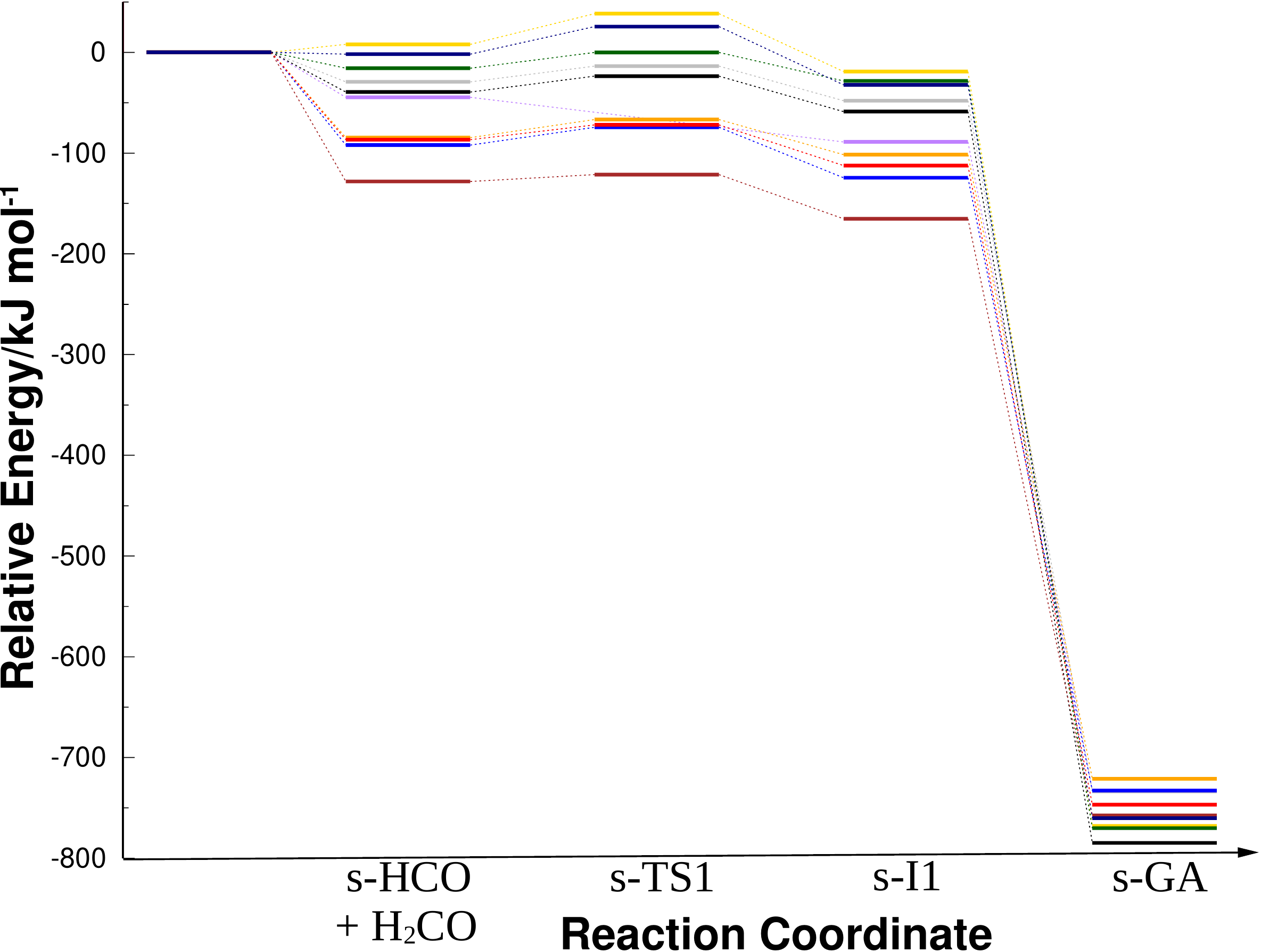}
    \caption{Formation of glycolaldehyde (GA) starting from an adsorbed HCO molecule.}
    \label{fig:allsurf2}
\end{figure}

The reaction path over all surfaces starting with H$_2$CO adsorbed are illustrated in Fig. \ref{fig:allsurf1}. This figure shows a tendency of energy decrease, initially due to the adsorption of the reactant molecules and followed by the chemical reaction itself, which presents more energetically stable products. This tendency is very similar to the gas phase reaction path, where the intermediate has a relative energy similar to the reactants and the final product presents very low energy. In contrast to the gas phase, in the catalysed route some reaction paths occur barrierlessly. The reaction paths starting with HCO adsorbed are illustrated in Fig. \ref{fig:allsurf2}. As previously  reported, the majority of cases did not lead to a decrease in energy barrier.

It is important to mention that the surface that presented the lower energy in the first step did not necessarily show the lower energy in the next step. In this way, it is challenging to point out a surface with a preferable topography for this chemical reaction to occur.

\subsubsection{Adsorption of Reactants and Intermediates}

It is important to know whether the newly formed products will be strongly bound to the surface and will remain on the grain for a long period, or if they are weakly bound and could be released to the gas phase using part of the energy released in the reaction. For this reason, we have also calculated the adsorption energy ($\rm E_{\rm adsorp}$) of each reactant, intermediate and products on each of the 10 model surfaces used in this work as

\begin{align}
    \rm E_{adsorp}=& \rm E_{total} -E_{surface} -E_{molecule} \label{eq:adsor}
    \end{align}
where $\rm E_{total}$ refers to the energy of the adsorbed configuration containing molecule and surface, while $\rm E_{surface}$ and  $\rm E_{molecule}$ correspond to the energy of the clean surface and that of the isolated molecule, respectively. The adsorption energies for each chemical species are shown in Table~\ref{tab:adsortab}.

\begin{table}
\centering
\caption{Adsorption energy of reactants, intermediates and product in kJ mol$^{-1}$. Eley-Rideal reaction type starting with H$_2$CO and HCO molecule adsorbed in the surface (s).}
\label{tab:adsortab}
\begin{tabular}{cccccc}
\hline
           & \multicolumn{2}{c}{Reagents} & \multicolumn{2}{c}{Intermediates} & Product \\
\# Surface & s-HCO        & s-H$_2$CO        & s-HCO          & s-H$_2$CO          & GA      \\ \hline
1          & -114        & -3          & -44          & -47             & -52     \\
2          & -84        & -25          & -3             & -60             & -53     \\
3          & -117          & -3      & -1             & -38           & -64     \\
4          & -26         & -24         & -39            & -59             & -52    \\
5          & -114         & -49         & -9            & -33             & -41     \\
6          & -84       & -35         & -43            & -41             & -75     \\
7          & -114        & -45          & -16            & -51             & -67     \\
8          & -151        & -49          & 38             & -7              & -100    \\
9          & -148        & -34          & -38            & -29             & -70     \\
10         & -113         & -39          & 68             & -4             & -73     \\ \hline
Averaged   & -106          & -31          & -8             & -37             & -64     \\ \hline
\end{tabular}
\end{table}


In Table 2, we would like to highlight the results obtained for the formaldehyde and glycolaldehyde species. The adsorption energy for formaldehyde ranges from -3 to -49 kJ mol$^{-1}$, and the average value is -31 kJ mol$^{-1}$. These results are in excellent agreement with experimental results found in the literature. According to an experimental study performed by \citet{noble2012desorption}, the adsorption energy of formaldehyde is estimated to be -27.1 kJ mol$^{-1}$, showing that our results present an average deviation of 4 kJ mol$^{-1}$ from the experimental one. It is possible to note that some surfaces, for example, surfaces 2 and 4, deviate in 2 and 3 kJ mol$^{-1}$ from the experimental result, and surfaces 6, 9, and 10 deviate from 12 kJ mol$^{-1}$, in the most.

For glycolaldehyde, we compare our results with the experiments performed by \citet{burke2015glycolaldehyde} on the adsorption energy of important COMs on amorphous solid water ices using infrared and temperature programmed desorption techniques. This experiment obtained a value of -46 kJ mol$^{-1}$ for glycolaldehyde, which can be compared with the results of this work shown in the last column of Table~\ref{tab:adsortab}. It can be seen that on average the deviation from the experiment was 18 kJ mol$^{-1}$. Several of our model surfaces showed deviations below 7 kJ mol$^{-1}$. 
Such deviations are expected for DFT calculations \citep{peverati2011improving,hammer1999improved}, and suggests that our method for generating the amorphous surfaces does not produce another significat source of error. 

The results in Table~\ref{tab:adsortab} also suggest that the interaction  between the reactants and the surface is strong enough to hold the reactants on the ice until the chemical reaction occurs. 

Let us first analyse the mechanism starting with H$_2$CO adsorbed (s-H$_2$CO), which as argued in the previous section is the preferred one. In this case the intermediate formed is seen to have an adsorption energy of -37 kJ mol$^{-1}$ on average. 
Recall that the formation of I1 is only slightly exothermic, and thus it would be formed on the grain with  low vibrational energy content, which could also be dissipated to the grain. Given the low temperatures of the ISM, it is likely that this intermediate would stay on the grain long enough for the second step of hydrogenation to occur, finally leading to glycolaldehyde. 

 As for the less favorable route, which starts with \ce{HCO} adsorbed and is attacked by a gas-phase H$_2$CO molecule, we see that the average adsorption energy of the intermediate is much lower in magnitude. In fact, there were two cases (out of ten) with positive adsorption energy, which would mean that the nascent intermediate would be promptly desorbed. This low magnitude of the adsorption energy gives a further indicative that this is not a viable mechanism, as the intermediate could in principle be desorbed right after formation, which would make it less prone for the subsquent step of hydrogen addition.

For the glycolaldehyde molecule, all surfaces presented negative adsorption energies  (Tab~\ref{tab:adsortab}), indicating  an attractive character to such molecule. However, the exothermicity of the second step of our proposed mechanism is large (on average releasing -751 kJ mol$^{-1}$), and for all of the ten cases, the product  glycolaldehyde molecule would in fact be formed with a vibrational energy content that is more than enough to overcome its adsorption energy and release it to the gas phase.

\section{Astrochemical Implications}

In this study, two sides of a problem were investigated and significant results were obtained that improved our understanding of the formation of the glycolaldehyde molecule.

In the gas phase side, results show that the carbon-carbon bond can be achieved through neutral-radical reactions with a low energetic barrier, enabling this process to occur in many astronomical environments where glycolaldehyde is found. However, this association would require a radioactive energy release (radioactive association) or  third-body collision for the stabilisation of the intermediate complex. In fact, \citet{woods2012formation} shows that the formation of glycolaldehyde in the gas phase is not sufficient to achieve the formation rates reported in the ISM.

\begin{figure}
	\includegraphics[width=\columnwidth]{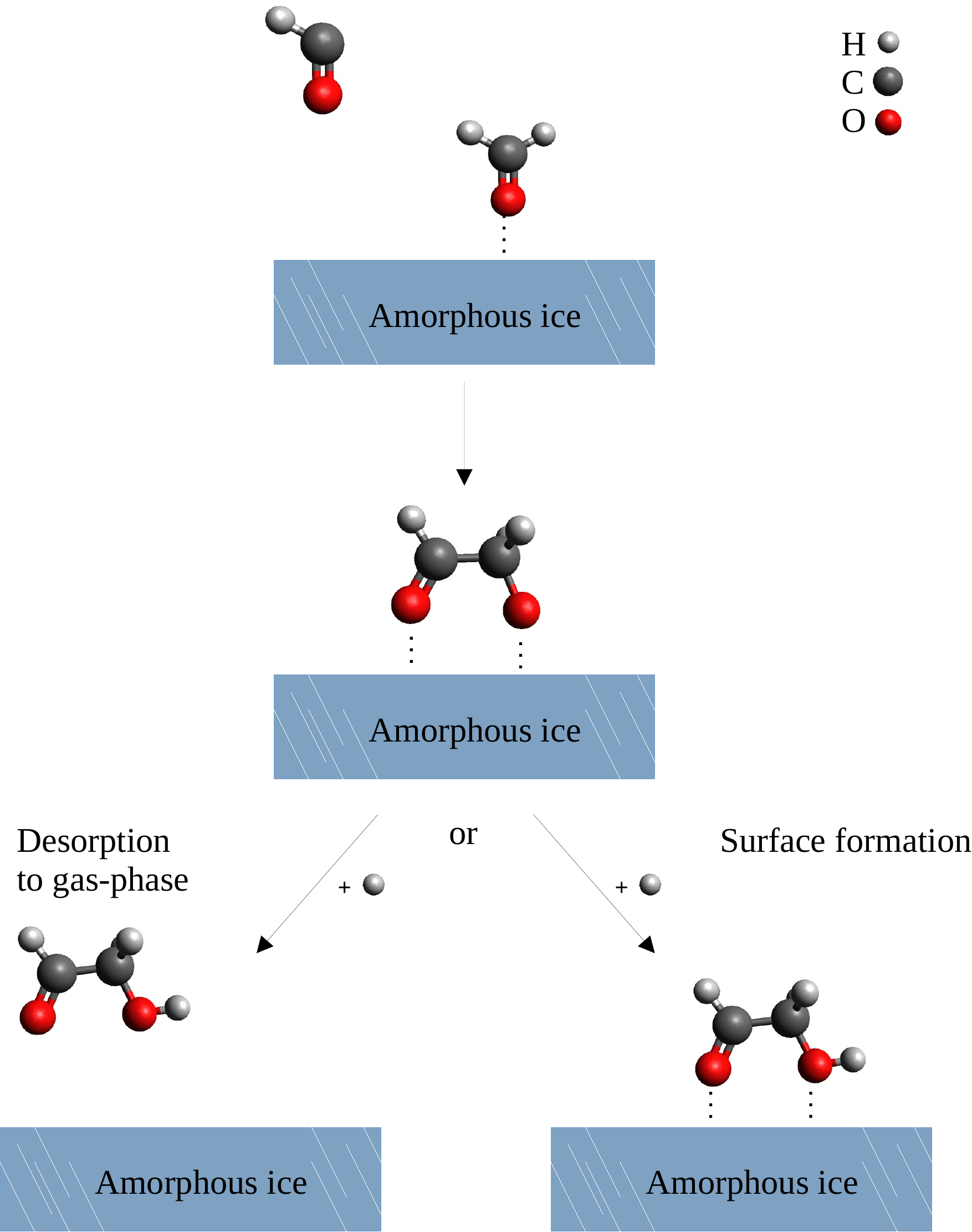}
    \caption{Most likely scenario of glycolaldehyde formation from the results obtained in this work. Dotted line means that the atom is bound to the surface through adsorption.}
    \label{fig:scenario}
\end{figure}

At the surface of a grain, the reactants proposed in our mechanism (formyl radical and formaldehyde) can be formed in the ISM in several ways. For example,
successive hydrogenation processes of CO can form the HCO and H$_2$CO, reactions already described in the KIDA database\footnote{Kinetic Database for Astrochemistry \url{https://kida.astrochem-tools.org/}} \citep{wak15}. Another process was recently described by \citet{molpeceres2021carbon}, wherein the interstellar amorphous water ice catalyses the two-step C + H$_2$O reaction to formaldehyde formation. The formaldehyde can be split in HCO radical and hydrogen atom in cosmic rays reactions, being a source of formation of this radical in the gas phase. Thus, our mechanism could be initiated also by a carbon atom on the amorphous ice, leading to our proposed reactants and culminating in the hydrogenation of the intermediate I1 on the ice surface as shown in Fig. \ref{fig:scenario}.

\begin{figure*}
	\includegraphics[scale=0.5]{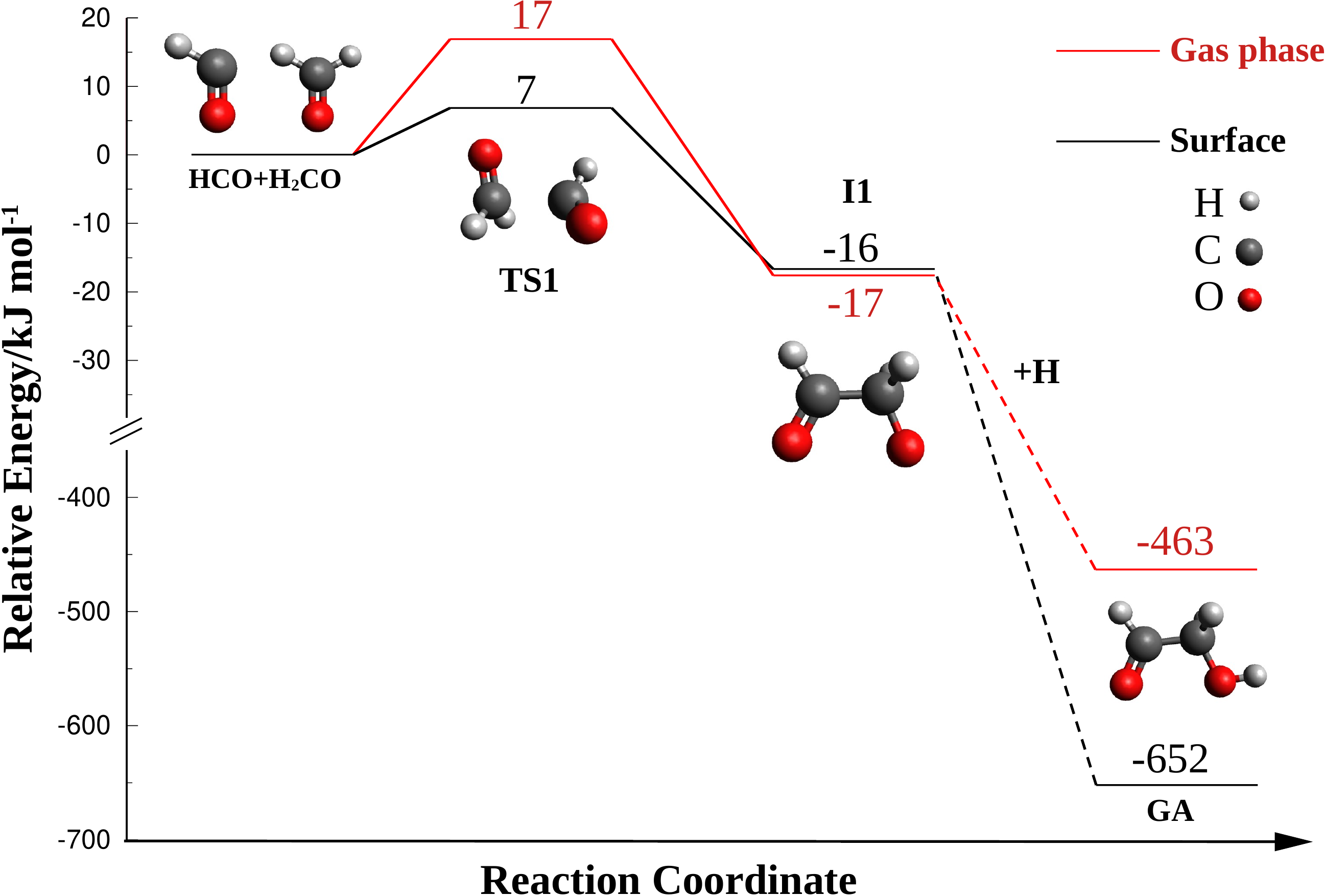}
    \caption{Comparison between the proposed mechanism in the gas phase and on the surface of amorphous ice. The displayed energy of the solid phase is the average of the energies of the ten surfaces studied. All energies are at DFT level.}
    \label{fig:dft-comp}
\end{figure*}

When we directly compare the relative energies between the mechanism in the gas phase and the same applied to the surface of amorphous ice, as can be seen in Fig. \ref{fig:dft-comp}, it is notable that the greatest influence occurs in the decrease of the energy barrier in the formation of the glycolaldehyde molecule.The energy barrier for this reaction reduces from a gas-phase value of 17 kJ mol$^{-1}$ (2040 K) to only 7 kJ mol$^{-1}$ (842K) on average for the surface. This barrier can be easily overcome by local heating induced by collisions or cosmic ray impacts, and the reactivity may be enhanced by quantum tunnelling \citep{paudel2014energy,bialy2020cold}. This leads to the question of how many other chemical reactions take place in the ISM in a similar fashion.

 In the case of the mechanism studied in this work, the exothermic chemical energy released in the formation of glycolaldehyde can lead to non-thermal surface desorption processes. With this desorption occurring glycolaldehyde would to released to the gas phase quickly after its formation. This hybrid mechanism between gas and solid phases would be the key to understanding the current abundances of glycolaldehyde in the ISM and other more complex COMs. Other reaction routes considering isomers of glycolaldehyde could also help to clarify this issue.

In comparison with other theoretical and experimental works,  \citet{Layssac2020} discussed a general formation route for sugars and polyol compounds. They investigated the formation of COMs, such as  glyceraldehyde and its saturated derivative glycerol, through VUV photolysis of interstellar ice analogues composed of H$_2$O and H$_2$CO. They proposed a formation mechanism for glycolaldehyde and polyoxomethylene (a formaldehyde polymer) based on a radical-non radical reaction between HCO and H$_2$CO. \citet{Ahmad2020} found that reactions between  two formaldehyde molecules exhibits a significant potential barrier, but due to quantum tunnelling these reactions can happen in the ISM. In addition to that, they also found that the chemical reaction is exothermic and capable of producing not only glycolaldehyde but also methyl formate and acetic acid. 

In a future study, it would be interesting to investigate the thermodynamic influence of some nonpolar amorphous surfaces with the presence of molecules such as CO and CO$_2$ in different proportions with water molecules. These studies should help quantify how much of the catalysis of chemical reactions would be directly linked to the surface composition.

It was shown in this work that small changes in the surface can lead to considerable changes in the final result. Therefore calculations with more details like explicit molecules, amorphous configuration and periodic conditions have a significant importance for correct results.

\section{Conclusions}

In this work we computationally  study the formation of glycolaldehyde from formaldehyde (H$_2$CO) and the formyl radical (HCO) over a two steps mechanism (reactions \ref{eq1} and \ref{eq2}),  occurring both in the gas phase and catalysed by a surface (amorphous water ice). 

In the gas phase, DFT and CCSD(T) calculations were used to construct the reaction path with the lowest possible energy. In the solid phase, periodic DFT calculations were used on a sample of ten amorphous ice surfaces in order to study their influence on the energy barrier of an Eley-Rideal mechanism. 
A consistent decrease in the reaction barrier was obtained in almost all of the different amorphous water surfaces explored with an average reduction of 49\% (which amounts to an average barrier of 7 kJ mol$^{-1}$) and reaching a decrease of 100\% in some cases. Even though each grain will have its unique amorphous structure and reactive sites, our results suggest  that the catalytic  effect is the predominant scenario. 

Given that the first step of the reaction is only slightly exothermic, our adsorption energy analysis indicates that the intermediate will remain on the grain for  a subsequent hydrogenation step, leading to glycolaldehyde. The large amount of energy released in this step will be greater than the adsorption energy of the product molecule and sufficient to overcome its adsorption energy. Therefore, the mechanism culminates in the release of glycolaldehyde to the gas-phase in a non-thermal desorption process. The results on the energy barrier, as well as the adsorption energies indicate that the Eley-Rideal mechanism with an adsorbed H$_2$CO being attacked by HCO is much more efficient then the analogous one starting with an adsorbed HCO.

The mechanisms involving the surface (the average value) lead to a product that, in addition to having a lower activation barrier than in the gas phase, is also much more exothermic than the product in the gas phase. Which may suggest that the formation of glycolaldehyde in cold regions can be much more efficient than in gas. The proposed mechanism and the computational framework proposed here may be transferred to the study of several other complex organic molecules. 

\section*{Acknowledgements}

The authors would like to thank the financial support provided  by the Coordena\c c\~ao de Aperfei\c coamento de Pessoal de N\'ivel Superior - Brasil (CAPES) - Finance Code 001, and
Conselho Nacional de Desenvolvimento Cient\'ifico e Tecnol\'ogico (CNPq), grants 311402/2021-6 and 311508-2021-9.

\section*{Data Availability}

The data of this paper will be shared on reasonable request to the corresponding author. The essential extra data can be seen at Supplementary Data section. 



\bibliographystyle{mnras}
\bibliography{reference} 








\bsp	
\label{lastpage}
\end{document}


 
 
\title{Supplementary Data - Glycolaldehyde Formation Mediated by Interstellar Amorphous Ice: A Computational Study}
\author{
M. A. M. Paiva,
S. Pilling,
E. Mendoza,
B. R. L. Galvão,
H. A. De Abreu\thanks{E-mail: heitorabreu@ufmg.br}}
 
\maketitle

\section{Analysis of exchange-correlation functionals for correct reproduction of TS1}

In order to state that the glycolaldehyde formation reaction is in fact catalysed when carried out on the surface of amorphous ice, we need a robust electronic structure method for predicting the energy barrier that would work both on the gas-phase and in the reaction catalyzed by the surface. However, when periodic condition calculations (without the surface and employing a large unit cell to simulate the gas-phase reaction) were performed using the PBE functional, the results revealed that there was no energy barrier to carbon-carbon bond formation in the gas-phase reaction. This contradicts the accurate CCSD(T)-F12/aug-cc-pVQZ results, which found the existence of an energy barrier.

To identify the failure in the reproduction of the energy barrier in the periodic condition calculations, a benchmark was carried out in which different methods were tested with ORCA and Quantum Espresso (QE) package programs. It is important to note that these calculations were performed in gas phase or in a unit cell containing only the molecules HCO and H$_2$CO.

\begin{figure}[!htb]
    \centering
	\includegraphics[scale=0.44]{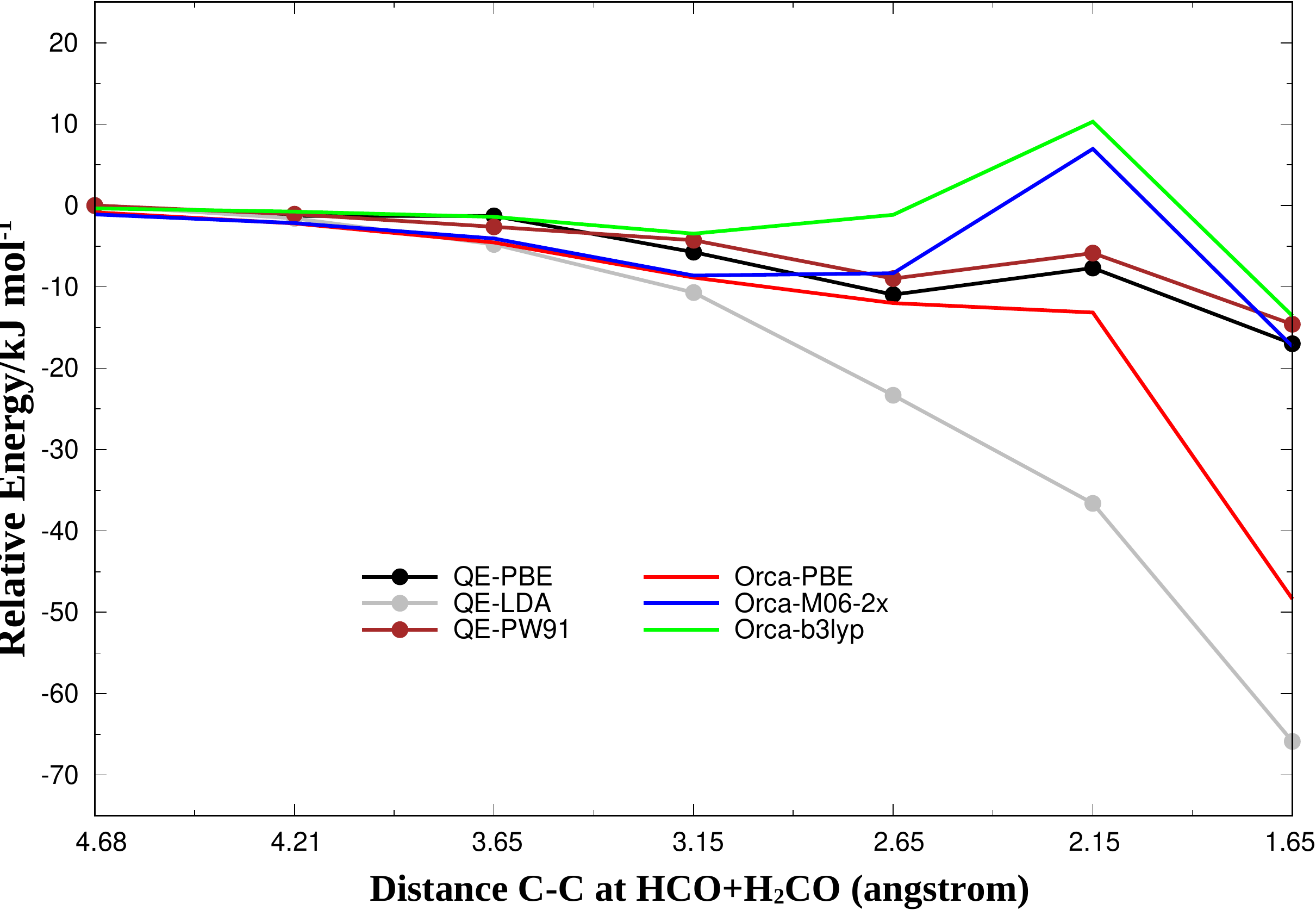}
    \caption{Energy profile obtained using different programs and methods to approximate the carbon-carbon coordinate between H$_2$CO+HCO molecules. The reference coordinates were optimized in the QE-PBE and used for point-point calculations in the other methods and/or programs. QE stands for Quantum Espresso, and the localised basis set used was aug-cc-pVTZ.}
    \label{fig:Supplementary-Data-fig1}
\end{figure}

As can be seen in Fig. \ref{fig:Supplementary-Data-fig1}, only the calculations in which hybrid-type exchange-correlation functionals were used managed to properly reproduce the existence of an energy barrier lying above the reactants. Thus, even calculations that use only localised bases and non-hybrid functional, as was the case with ORCA PBE, failed to describe the energy barrier, which shows that the deficiency for the description of the energy barrier lies in the choice of the inappropriate functional and not
the use of plane waves.

\begin{figure}[!htb]
    \centering
	\includegraphics[scale=0.44]{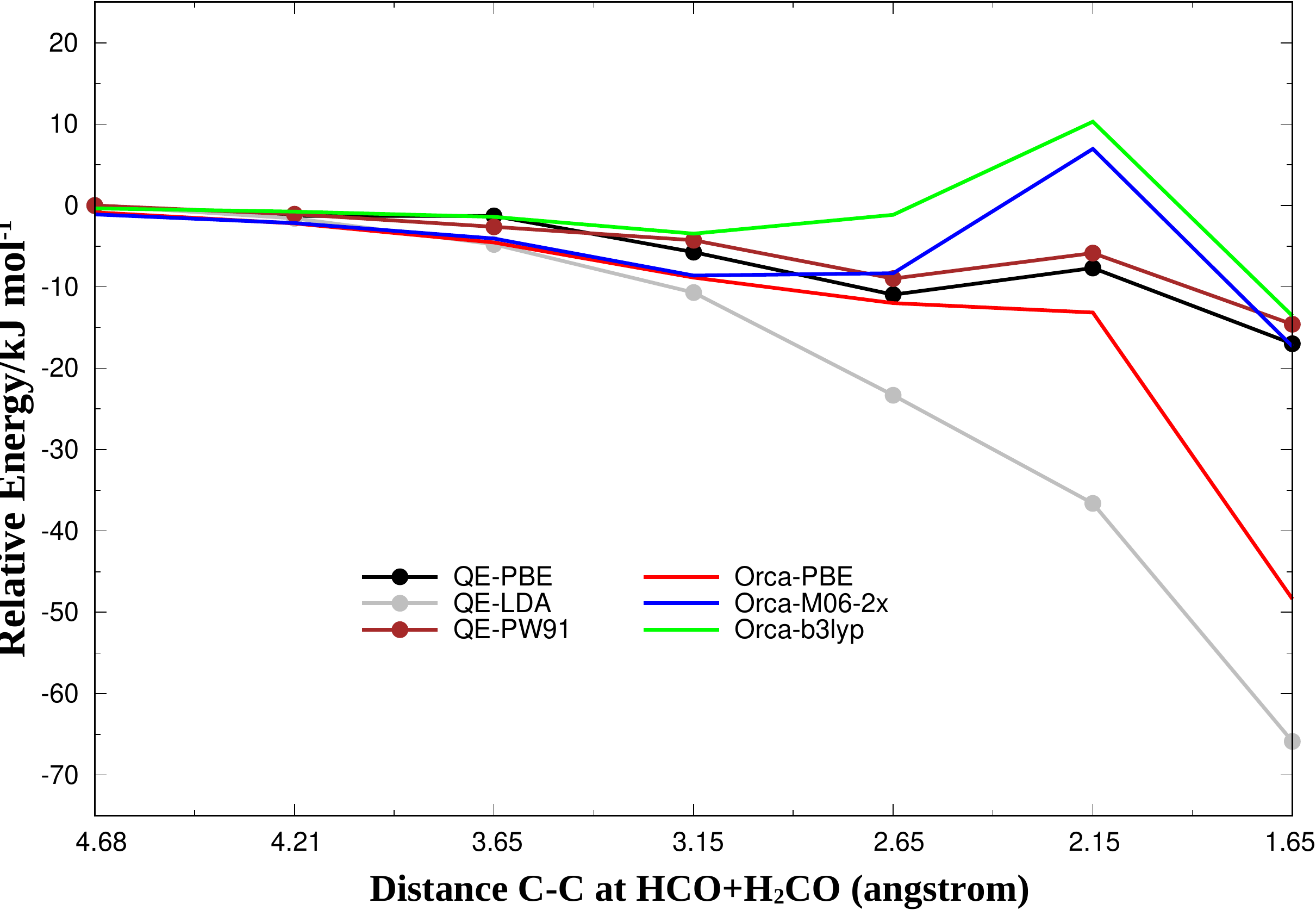}
    \caption{Energy profile obtained using different programs and methods to approximate the carbon-carbon coordinate between the H$_2$CO+HCO molecules. The reference coordinates were optimized in ORCA-M06-2x and used for single-point calculations in other methods and/or programs. The localised basis set used was aug-cc-pVTZ.}
    \label{fig:benchmark2}
\end{figure}

In Fig. \ref{fig:Supplementary-Data-fig1} the HCO and H$_2$CO approximation geometries were obtained under PBE XC-functional using the Quantum Espresso program. Comparatively, the geometries at each point in Fig. \ref{fig:benchmark2} were obtained using the ORCA program package and M06-2X XC-functional optimizations.

Fig. \ref{fig:benchmark2} confirms that only calculations with hybrid functionals reproduce the energy barrier. The only exception occurs for  HSE functional, which partially reproduced the energy barrier, all other calculations with hybrid functionals obtained very close energies, regardless of the chosen basis set or program. It can be  seen that B3LYP calculations from both ORCA and Quantum Espresso present almost identical energies in the correct description of the system.

This benchmark was necessary so that when comparing the results of the energy barrier in the gas phase and on the ice surface, there is no error or bias associated with the functional or basis set used.

\section{Formation Route H$_2$CO + HCO + H}

\begin{table}[htpb]
\centering
\caption{DFT, Coupled Cluster and zero-point energy for each structure. Energy values relative to the reactant, except ZPE which is in terms of total energy. All values are in kJ mol$^{-1}$ except T1 diagnostic which is in SI.}
\label{tab:a1a}
\begin{tabular}{ccccc}
\hline
\textbf{Name} & \textbf{M06-2x + ZPE} & \textbf{CCSD(T)-F12 + ZPE} & \textbf{ZPE} &  \textbf{T1 diagnostic} \\ \hline
H$_2$CO+HCO+H & 0 & 0 & 105.91 & 0.01477 0.02031 \\
TS1 & 16.86 & 27.23 & 115.67 & 0.03055 \\
I1 & -17.62 & -5.70 & 125.34 & 0.02221 \\
TS2 & 41.60 & 57.62 & 104.44 & 0.03361 \\
I2 & -82.71 & -77.50 & 128.57 & 0.01722 \\
GA & -463.42 & -455.64 & 162.99 & 0.01388 \\ \hline

\end{tabular}
\end{table}

\begin{longtable}{ccc}
\label{tab:a1b}\\
\hline
\textbf{Name} &
  \textbf{Vibrational Frequencies (cm$^{-1}$)} &
  \textbf{Cartesian coordinates (XYZ)} \\ \hline
\endhead
H$_2$CO+HCO &
  \begin{tabular}[c]{@{}c@{}}1108.19\\ 2004.66\\ 2723.22\\ -----//------\\ 1215.95\\ 1280.80\\ 1541.63\\ 1875.47\\ 2942.07 \\ 3015.19\end{tabular} &
  \begin{tabular}[c]{@{}c@{}}C     -1.339035    0.469623   -0.001171\\ O     -0.536175   -0.375263    0.001671\\ H     -1.580932    1.096035   -0.896329\\ -----//------\\ C     -1.358419    0.489508   -0.000007\\ O     -0.535656   -0.375118    0.000013\\ H     -1.569442    1.086451   -0.902319\\ H     -1.944151    0.729853    0.902313\end{tabular} \\ \hline
TS1 &
  \begin{tabular}[c]{@{}c@{}} 322.29i\\ 39.31 \\ 215.40 \\ 270.50 \\ 613.13 \\ 667.97 \\ 1117.86 \\ 1136.92 \\ 1249.13 \\ 1492.47 \\ 1663.50 \\ 1999.29 \\ 2900.96 \\ 2942.08 \\ 3013.59\end{tabular} &
  \begin{tabular}[c]{@{}c@{}}C      0.015694    0.004595   -0.000980\\ C      2.123542    0.001983    0.010076\\ O     -0.147348    1.219111    0.001039\\ H     -0.044468   -0.583738    0.926795\\ H     -0.048414   -0.578081   -0.934061\\ O      2.733353   -0.453136    0.886385\\ H      2.428928    0.773048   -0.722912\end{tabular} \\ \hline
I1 &
  \begin{tabular}[c]{@{}c@{}}121.86 \\ 344.16 \\ 532.10 \\ 703.09 \\ 1019.36 \\ 1022.62 \\ 1148.10 \\ 1156.87 \\ 1312.66 \\ 1375.93 \\ 1400.93 \\ 1882.69 \\ 2927.44 \\ 2972.20 \\ 3046.23\end{tabular} &
  \begin{tabular}[c]{@{}c@{}}C     -1.330934    0.001337   -0.014427\\ C      0.190680    0.023147    0.041607\\ O     -1.852920    1.115671   -0.579034\\ O      0.846034   -0.972703    0.034203\\ H     -1.664192    0.016612    1.040764\\ H     -1.696687   -0.937150   -0.445702\\ H      0.637619    1.032256    0.088928\end{tabular} \\ \hline
TS2 &
  \begin{tabular}[c]{@{}c@{}} 2138.29i\\ 104.57 \\ 134.98 \\ 195.40 \\ 407.99 \\ 547.27 \\ 703.87 \\ 991.42 \\ 1242.57 \\ 1432.20 \\ 1512.40 \\ 1672.80 \\ 2048.31 \\ 3055.32 \\ 3158.15\end{tabular} &
  \begin{tabular}[c]{@{}c@{}}C     -1.084073   -0.151421   -0.037989\\ C      1.844694    0.301645    0.046127\\ O     -0.535186    0.933646   -0.317649\\ H     -1.455639   -0.340978    0.968994\\ H     -1.206873   -0.926067   -0.795008\\ O      1.674162   -0.779552    0.389545\\ H      0.762914    0.962726   -0.254019\end{tabular} \\ \hline
I2 &
  \begin{tabular}[c]{@{}c@{}}138.59 \\ 182.98 \\ 341.42 \\ 529.78 \\ 852.03 \\ 859.96 \\ 1139.19 \\ 1211.78 \\ 1308.92 \\ 1392.35 \\ 1476.51 \\ 2013.99 \\ 3027.75 \\ 3110.26 \\ 3867.57\end{tabular} &
  \begin{tabular}[c]{@{}c@{}}C     -0.408151   -0.146169    0.000933\\ C      1.093915   -0.193081    0.211099\\ O     -0.844704    1.105178   -0.435910\\ H     -0.867895   -0.407539    0.957144\\ H     -0.640481   -0.930678   -0.725039\\ O      1.742990   -1.084675    0.591295\\ H     -0.075674    1.656963   -0.599522\end{tabular} \\ \hline
GA &
  \begin{tabular}[c]{@{}c@{}}216.42 \\ 311.18 \\ 393.27 \\ 735.34 \\ 774.71 \\ 892.94 \\ 1120.11 \\ 1166.13 \\ 1266.44 \\ 1309.26 \\ 1402.57 \\ 1449.89 \\ 1488.54 \\ 1858.64 \\ 2990.67 \\ 3021.36 \\ 3041.69 \\ 3787.48\end{tabular} &
  \begin{tabular}[c]{@{}c@{}}C     -0.626413   -0.118125   -0.018700\\ C      0.810526   -0.503185    0.183238\\ O     -0.759090    1.226322   -0.346352\\ O      1.695739    0.296918    0.063630\\ H     -1.168900   -0.360209    0.903030\\ H     -1.037197   -0.763754   -0.804060\\ H      1.017409   -1.553088    0.446035\\ H      0.129647    1.603039   -0.365813\end{tabular} \\ \hline
  
\end{longtable}

\newpage

\section{Surface \#1}
Tetrahedral cell $\alpha=\beta=\gamma=90^o$
\begin{verbatim}
    CELL_PARAMETERS (angstrom)
 10.0   0.0   0.0
  0.0   10.0   0.0
  0.0   0.0   20.0

ATOMIC_POSITIONS {angstrom}
O        4.404634248   2.821291111   4.145276240
H        4.540842279   2.701425629   3.153528929
H        5.122711130   3.415991166   4.480059914
O        7.875884872   1.896459919   5.781581559
H        8.676628274   1.715544590   5.234465305
H        7.119818761   1.317539030   5.490345387
O        7.072462388   6.122430219   1.666492504
H        7.584024807   5.227975176   1.574549252
H        7.530764678   6.865389662   1.205527417
O        0.676184969   2.449419890  -0.493572554
H        0.833338423   3.087011303   0.247379443
H        1.378258674   1.772348912  -0.364641221
O        6.836062737   6.492421491   4.251915165
H        7.006798725   6.327344595   3.262803719
H        6.963560052   5.622406199   4.717816039
O        1.725669252   6.411999283   4.978318694
H        2.670269005   6.533425746   4.694297598
H        1.603582714   5.439863059   4.811877470
O        1.843206263   3.790215692   4.185788226
H        2.795967563   3.494402366   4.219639778
H        1.317544031   3.021543197   4.534569801
O        6.455678487  10.024301020   2.242269674
H        6.931178285   9.878808168   3.092116252
H        5.558237892   9.596324508   2.400204177
O        4.737419099   2.415395903   1.508676021
H        5.548629672   1.860471579   1.530226291
H        4.039487622   1.830632467   1.104330248
O        4.198393268   6.512530390   3.845997729
H        5.158538104   6.493780377   4.136692469
H        4.189942469   5.979981011   3.000206293
O        3.155760706   7.739586069   0.398540658
H        3.657913008   6.890233036   0.452393491
H        2.210192481   7.460030659   0.532822679
O        6.542159820   4.118747128   5.500314730
H        7.170552396   3.319198172   5.586568011
H        6.232907604   4.295710428   6.405022012
O        1.729449415   0.115543420   3.047959963
H        1.181022098  -0.731591124   3.127093390
H        2.670476176  -0.229602535   3.117927263
O        0.635896213   6.869462700   1.011050415
H       -0.142856391   6.976371665   0.435924867
H        0.387724017   7.260367303   1.914505356
O        0.523635015   1.507170093   4.928163170
H        1.017800043   0.961391782   4.225641552
H        0.822113567   1.160173459   5.785798750
O        5.524973491   0.604274234   5.240219947
H        5.050849550   0.381804677   6.058413229
H        5.010942616   1.360528076   4.826921440
O        4.504768745   5.388046983   1.351117088
H        5.470372707   5.654691463   1.334003678
H        4.505795389   4.416077855   1.197299745
O        8.093943452   8.607382555   0.678124697
H        7.422902870   9.115406656   1.238973817
H        8.949074706   8.996083279   0.933183292
O        8.147105847   3.816720246   1.775865747
H        9.138892580   3.897628727   1.825769610
H        7.990439627   3.122823871   1.070742461
O        0.855894993   4.180854502   1.679954646
H        0.896076006   5.141481804   1.445013025
H        1.300893448   4.058887151   2.565036954
O        4.062182221   8.894355899   2.650649574
H        4.156190865   8.070238347   3.217819356
H        3.732498291   8.505669559   1.780602664
O        0.327438520   7.902653997   3.386918305
H        0.832378315   7.293836558   4.045098870
H       -0.519094996   8.209121496   3.821943704
O        2.700496539   0.749978803   0.514609431
H        2.215213492   0.537328428   1.357984986
H        2.973074184  -0.120197234   0.155863625
O       -1.957283808   1.972809035  -0.205963789
H       -0.972359249   2.025440435  -0.397653626
H       -2.158165797   1.029395327  -0.076133593
O        8.024047504   8.821054212   4.416526997
H        7.527728729   7.945013868   4.503397539
H        8.008234693   9.241976342   5.292618258
\end{verbatim}

\newpage

\section*{Surface \#2}
Tetrahedral cell $\alpha=\beta=\gamma=90^o$
\begin{verbatim}
    CELL_PARAMETERS (angstrom)
 10.0   0.0   0.0
  0.0   10.0   0.0
  0.0   0.0   20.0

ATOMIC_POSITIONS {angstrom}
O             8.3398218135        7.6406247780        5.4866150529
H             7.6948424415        8.2073757286        4.9376234814
H             8.6889566399        8.2502605736        6.1608428418
O             0.9981259055       -0.2530338873        5.0789124626
H             0.3854107912        0.3549032660        4.5753739781
H             0.9539120586       -1.1492456416        4.6530871708
O             5.6557809600        1.5797844915        3.8343631498
H             4.7826843119        1.4059072348        4.2595361848
H             5.4655013680        1.9500238973        2.9228789032
O             2.3338909711        5.2236732929        5.2144915292
H             3.3328386516        5.3742116050        5.0586118571
H             2.2352295165        5.0081377251        6.1573724849
O             4.8363751876        5.7309184448        4.7454779977
H             5.2602100632        5.5789262161        3.8492308299
H             5.5911529507        5.5336008563        5.4036365801
O             2.6296632070        2.6449474780        1.0063379589
H             2.4462631406        2.7363538092        1.9773590229
H             2.7654285496        1.6616712616        0.8310438002
O             6.3191909316        5.0875192553        2.5779612884
H             6.6446857055        5.7050691872        1.8776875189
H             5.8700462040        4.3336849780        2.0871596148
O             7.6316000774        2.4228820187        0.1256496219
H             7.9254402799        1.5763469052        0.6260226819
H             7.6554179819        2.1902177795       -0.8175020560
O             7.6358826436        3.3736834049        4.3359066074
H             7.3896416154        4.0208934750        3.6208874454
H             6.8705406926        2.7315219999        4.3237828059
O             0.6568665379        7.1355781575        4.1782171333
H             1.2289250581        6.4780247978        4.6565462682
H            -0.2284050321        7.1696309985        4.6407827488
O             2.9023563160        7.4703562043        0.4932770938
H             2.1848474912        6.9707664170        0.9537977459
H             3.6401202590        7.5661187788        1.1372406023
O             6.5255926747        8.9915844371        4.1556816634
H             6.4780572408        9.9690210990        3.9870409327
H             5.6978988323        8.7948488353        4.7006535274
O             2.0246133282        2.9701750730        3.6819303214
H             2.5269917487        2.3625971458        4.2887247322
H             2.0433272207        3.8452463366        4.1514296778
O             6.9395340866        5.2820421742        6.2520893732
H             7.4817275215        6.0889120142        6.0726298479
H             7.3362777537        4.5624616086        5.6946278279
O             4.3323107064        8.2574268297        5.4438799322
H             3.5798119816        8.6335901502        4.9559129393
H             4.4334812344        7.3115495660        5.1266429673
O            -0.6796610322        8.1043811063        0.2722566602
H            -0.1642533943        8.2797185821       -0.5335105271
H            -0.1290409206        7.4151177146        0.8061766094
O             0.1516420699        3.9651216900        0.5699263444
H             0.9458182939        3.3742797648        0.5871841698
H            -0.6595040238        3.4156442338        0.4952402467
O             9.6579767160        1.6383162957        3.6400697149
H             8.9383279225        2.2687519475        3.9281905066
H            10.4740657331        2.2073406938        3.5588035687
O             3.3762280840       10.1698585931        0.4559640848
H             4.1918308134        9.8654011241        0.9209960422
H             2.9015138336        9.3064502868        0.3446424979
O             5.2397804607        8.4771284969        1.7179431522
H             5.8694164768        7.9778124316        1.1440369734
H             5.7399800203        8.6238945566        2.5603678492
O             3.2130499308        1.0746104262        5.2203029939
H             3.4438825013        1.1415135268        6.1621472828
H             2.3658531936        0.4920735693        5.1792064400
O             8.3751423099        0.4061837041        1.5640517600
H             8.8137708808       -0.3815957912        1.1594723555
H             8.9725146389        0.7803251180        2.2657601354
O             7.0042727540        6.7758100802        0.4039880229
H             6.7349827370        6.3522836169       -0.4277166072
H             7.8776117307        7.2415432103        0.2316057722
O             0.6326420331        6.3388826038        1.5922597480
H             0.5645643540        6.4935895513        2.5738272764
H             0.3833857051        5.3975285056        1.3295049870
O             5.2937944971        2.8946923990        1.4855792534
H             4.3595004455        2.8597575001        1.1194457925
H             5.9701460650        2.6503312023        0.7989303763
\end{verbatim}

\newpage

\section*{Surface \#3}
Tetrahedral cell $\alpha=\beta=\gamma=90^o$
\begin{verbatim}
    CELL_PARAMETERS (angstrom)
 10.0   0.0   0.0
  0.0   10.0   0.0
  0.0   0.0   20.0

ATOMIC_POSITIONS {angstrom}
O        7.136085269   4.225947122   1.736249778
H        7.892520797   3.661260471   1.316514243
H        6.814411036   4.835978284   1.039875138
O       -0.986176034   2.756697178   0.803535380
H       -1.158834669   1.806781218   0.990945266
H       -0.208510488   3.014590537   1.367670017
O        1.669357713   5.547297441   6.232590417
H        2.591562696   5.222359113   6.343572523
H        1.713011470   6.250283963   5.533625524
O        2.975430787   1.950542167   5.379325836
H        3.006120236   1.067811195   4.954986443
H        2.027491132   2.247911456   5.363715560
O        6.057325064   1.431197622   5.322956034
H        6.450351641   0.543960193   5.087505089
H        5.185395818   1.257421238   5.724459059
O        5.548458515   2.208817047   2.718930660
H        5.763681434   2.058374823   3.675389527
H        6.151670110   2.913206115   2.352706586
O        6.265470830   6.780323925   0.467191696
H        6.162190403   7.719312081   0.122613087
H        5.408473492   6.632919933   0.949537792
O        3.973751722   0.121797912   1.892082035
H        3.018364875   0.390964583   1.813424690
H        4.491255323   0.943451874   2.100263108
O        4.916969972   6.608863413   4.389964308
H        4.521368000   5.881409872   4.935805024
H        5.841942250   6.246008280   4.227164081
O        7.305323960   5.475947309   4.054793876
H        8.085803644   6.026252561   4.310362205
H        7.415448741   5.052436005   3.144124272
O        8.533455129   7.381109234   1.639364596
H        8.345015292   7.286895662   2.590430524
H        7.664923347   7.072358560   1.195129914
O        0.942610528   3.622962167   2.531119612
H        0.778810577   3.552833578   3.523689071
H        1.956763505   3.596705442   2.356763626
O        7.144438921   9.167782106   4.317063374
H        6.390299096   8.571287219   4.152537358
H        7.831859871   8.593338961   4.763710460
O        3.943704148   6.308150880   1.865291682
H        4.160721423   6.485370028   2.822871634
H        3.002118542   6.553699507   1.713665330
O        8.499933077  10.092831694   1.773174674
H        8.007457346  10.096737976   2.622213672
H        8.567973809   9.115634042   1.570815868
O        0.576848748   3.344082041   5.170593337
H        0.913200344   4.186269249   5.608054693
H       -0.372209867   3.245140264   5.468847143
O        3.403279129   3.633334225   1.882029631
H        3.710450099   4.576438457   1.833443657
H        4.122783228   3.120828269   2.332004807
O        5.965074317  -0.585751274  -0.000470008
H        6.702112114  -0.123028349   0.446414757
H        5.166310393  -0.362022739   0.538502115
O        3.555515571   9.150443394   4.354434167
H        4.115477743   8.360079049   4.532015512
H        3.762763400   9.431612209   3.408481450
O        1.401817711   7.505998351   4.302571283
H        2.086598430   8.237751171   4.280359076
H        1.336737822   7.089840344   3.385932031
O        1.084981154   6.302248707   1.961150127
H        0.228813632   6.642871886   1.600139340
H        0.940852593   5.324619761   2.126486803
O        4.211060795   4.332179713   5.857811151
H        4.945316513   4.137095924   6.466322871
H        3.862820111   3.442792517   5.561880198
O        7.934203249   3.436856852   5.891339696
H        7.608394917   4.105002779   5.239986791
H        7.313268080   2.675943136   5.809328684
O        1.344042393   0.741843389   2.379018849
H        1.204583522   1.715863905   2.423520565
H        0.457842268   0.389514297   2.142661779
O        8.982328531   7.380771422   5.172089023
H        9.942897742   7.548701170   4.860418401
H        9.062127805   7.072299132   6.092756399
\end{verbatim}

\newpage

\section*{Surface \#4}
Tetrahedral cell $\alpha=\beta=\gamma=90^o$
\begin{verbatim}
    CELL_PARAMETERS (angstrom)
 10.0   0.0   0.0
  0.0   10.0   0.0
  0.0   0.0   20.0

ATOMIC_POSITIONS {angstrom}
O        7.524053336   1.509989555   0.239401540
H        6.907488730   1.143957234  -0.413271708
H        7.348509983   2.482528787   0.270726988
O        1.132748263   4.496325281   3.148913856
H        1.459610260   4.327765776   4.105653746
H        1.919337743   4.739424990   2.582007175
O        9.376219518   3.009363036   6.129581096
H        8.771520630   3.680456532   5.748706677
H        9.024593538   2.127109729   5.880156411
O        4.857681533   8.399535579   1.520072539
H        4.657224251   7.888505849   2.362808735
H        4.134080231   8.122711187   0.913082488
O        7.127012153   0.440646597   2.783401355
H        6.684920252  -0.388149352   2.530441938
H        7.330434584   0.880464855   1.916067590
O        7.291231285   4.097397666   1.082384960
H        7.313513580   5.080760374   0.908571323
H        8.072609381   3.855815719   1.632446499
O        4.921767206   0.801873956   4.642522097
H        4.247809895   0.720488140   3.919349638
H        5.731988763   1.017895124   4.130860842
O        1.745515557   4.253535606   5.648800154
H        0.962714782   3.720973186   5.959194807
H        2.585509264   3.743993418   5.879580420
O        3.192478960   5.304412639   1.672969718
H        3.552904921   5.918200316   2.395126946
H        2.920513924   5.914846179   0.943075936
O        4.189813983   6.833638907   3.577648593
H        3.529995962   6.974593012   4.337241367
H        4.806598372   6.117880054   3.910939731
O        3.572366505   0.795208374   2.232635365
H        4.055771716   1.635910750   2.006244006
H        4.097204254   0.035099906   1.876572140
O        6.209245107   8.774003734   5.964252192
H        5.567974141   9.379380487   5.507365617
H        5.747892239   8.484920329   6.766605606
O        8.413019847   0.344978693   5.299644820
H        7.700204538  -0.168643570   5.758379025
H        8.084353599   0.415925118   4.370754323
O        4.099105796   3.126089663   6.132633950
H        4.666917782   3.737551020   5.601791196
H        4.309659239   2.247171681   5.744726072
O        7.975625789   5.299379100   5.069564930
H        7.760509786   5.934150499   5.773067451
H        8.516404919   5.824209988   4.397783337
O        7.221902585   6.784023335   0.552191462
H        8.041493396   7.313769920   0.645263784
H        6.487343609   7.333245700   0.902984573
O       -0.416414760   2.737839793   2.017362018
H        0.230158213   3.390688107   2.453066748
H       -0.181392936   2.711939868   1.074383884
O        5.510934206   4.640416702   4.314988286
H        5.401681450   4.038639034   3.536578689
H        6.481712772   4.792573162   4.487852765
O        2.502083859   7.007751557   5.588494033
H        1.765504502   7.637106543   5.409431220
H        2.108612652   6.115126421   5.738768774
O        4.860382929   3.213222307   1.980180221
H        4.242217692   3.944632082   1.708826242
H        5.735692802   3.460722908   1.581737549
O        2.443902510   7.421019744   0.103489592
H        2.390485496   7.448800847  -0.864136949
H        1.507819130   7.626373691   0.434189526
O       -0.007240408   7.974880861   1.048179551
H       -0.177888676   7.426172044   1.867216228
H        0.183029305   8.881811411   1.406826680
O        0.871447623   0.268833924   2.347392880
H        1.823537098   0.529253946   2.228500099
H        0.351937063   1.113360284   2.338925839
O       -0.457529367   6.604169388   3.360504951
H       -0.072297011   7.359098199   3.900829832
H        0.228075073   5.866722582   3.311772230
O        0.527282769   8.758020652   4.629727848
H        0.776078807   9.331609019   3.856173551
H       -0.212079911   9.279409241   5.050625105
\end{verbatim}

\newpage

\section*{Surface \#5}
Tetrahedral cell $\alpha=\beta=\gamma=90^o$
\begin{verbatim}
    CELL_PARAMETERS (angstrom)
 10.0   0.0   0.0
  0.0   10.0   0.0
  0.0   0.0   20.0

ATOMIC_POSITIONS {angstrom}
O        0.772300263   7.342347745   2.216936547
H       -0.192706033   7.275901568   1.929791172
H        0.954202035   6.733418706   2.976948890
O        2.812727444   1.043026235   3.044329195
H        1.968372652   0.514473514   3.014951382
H        2.640404371   1.881289834   2.518359120
O        7.185506000   5.893396378  -0.085252162
H        6.460304603   6.045966947  -0.712549109
H        7.002728070   5.029038683   0.374293654
O        8.490712480   0.355077016   0.451934046
H        8.523198851  -0.562550128   0.831851288
H        8.939968434   0.972860026   1.069478476
O        6.582326676   8.088291631   3.574359765
H        7.051007416   7.865211315   4.428599177
H        6.210902545   9.018757579   3.606936336
O        4.867313427   4.148682507   3.396823658
H        5.562602356   3.989359339   2.702631428
H        4.710507622   5.142896838   3.403712268
O        5.648300146   2.392789018   5.245663949
H        4.836578491   2.103675869   5.718825880
H        5.350526216   3.153561849   4.657449278
O        3.229888475   7.300380836   0.906979875
H        3.514232955   8.145660359   0.445436713
H        2.281836990   7.397130765   1.178012802
O        7.801920123   7.094605228   5.755378432
H        7.691251350   6.132060859   5.683404578
H        8.758520460   7.231850087   5.999622459
O        0.495799316  10.243861228   5.979831205
H        0.544727351   9.955305704   5.025268044
H        0.445308093   9.379115052   6.449926198
O        5.976632029   1.147074530   0.755420805
H        6.904498225   0.794279088   0.554488363
H        5.818742045   0.897420807   1.712563748
O        0.526689135   9.682822341   3.328368605
H        0.600958915   8.761561746   2.896251600
H       -0.312697596  10.083796362   3.040173164
O        4.080861541  -0.485623658  -0.369825828
H        4.771371494   0.132447569   0.007022641
H        3.418841725   0.086467178  -0.790459374
O       -0.761128365   2.549534851   2.317207412
H       -0.987232877   2.328571719   3.266202235
H        0.119587679   2.980023722   2.322660216
O        6.763733464   3.624410781   1.408821438
H        6.430094876   2.741770181   1.047100579
H        7.675706306   3.403574786   1.735763277
O        5.407970282   0.464736935   3.314854753
H        5.599505557   1.153221212   4.013215996
H        4.413875316   0.508552448   3.201596389
O        2.357584264   4.331937122   6.195051179
H        3.158817934   4.578609788   6.686423335
H        2.484186004   3.356244044   5.966455826
O       10.432013386   7.465084037   6.463697698
H       10.658392619   6.931554546   7.244638644
H       11.040889676   7.122245263   5.727740374
O        1.877779500   6.350688921   4.544205593
H        2.778876131   6.631862344   4.217841236
H        2.042555858   5.485674044   5.030284477
O        8.273475632   7.762041421   1.496527184
H        7.637436225   7.860223482   2.266132657
H        7.837095030   7.117323547   0.865630903
O        8.392409414   1.647724560   4.712478650
H        8.964512122   1.246992805   5.408887230
H        7.531697183   1.901995720   5.112420630
O        2.810438086   1.769201726   5.668557682
H        2.810758863   1.551588192   4.689708855
H        2.061831849   1.216861528   6.018034023
O        2.345368967   3.481643614   2.109224359
H        3.056766662   3.786075443   2.724588068
H        2.675814296   3.849799409   1.240851893
O        4.272109851   6.746930732   3.318108929
H        3.960015340   6.993450796   2.394523620
H        5.119056548   7.263521159   3.469738126
O        3.697458883   4.721798030   0.057770665
H        4.599584187   4.526478570   0.367216876
H        3.552890225   5.685829503   0.266396640
\end{verbatim}

\newpage

\section*{Surface \#6}
Tetrahedral cell $\alpha=\beta=\gamma=90^o$
\begin{verbatim}
    CELL_PARAMETERS (angstrom)
 10.0   0.0   0.0
  0.0   10.0   0.0
  0.0   0.0   20.0

ATOMIC_POSITIONS {angstrom}
O             0.2995472385        2.2545384206        1.7715146398
H             1.1776301470        1.9281188739        1.3575176451
H             0.1304605572        1.5589075970        2.4685520055
O             0.6385868033        6.1383326068        4.7746914880
H             0.8783410483        5.4114429796        4.1084046955
H             1.4579702553        6.7272809915        4.7775909682
O             4.7947731236        2.3550821506        1.9555906708
H             5.4670022510        1.7530275136        2.4075823366
H             5.3620279318        3.0926866819        1.5598989435
O             5.3853209390        4.0933356315        5.4334119749
H             5.2510538004        4.9532972264        4.9394072678
H             4.6161270319        3.5478896316        5.1085804327
O             6.5124197804        0.8938868795        3.4216070328
H             6.8523222722        0.0831238753        2.9010881381
H             7.2518755914        1.4847185467        3.7234553668
O             5.1232447350        6.6419674310        1.4553070579
H             5.5967353714        5.8471600102        1.0754086592
H             4.1610568052        6.4941708908        1.2158901675
O             0.4845063801        0.1047910346        3.5055385997
H             1.1952426551        0.3026899220        4.1627391084
H             0.9703428178       -0.4420692897        2.8016070868
O             1.4397650128        4.3442670379        3.0242512459
H             2.1380497718        3.7947787407        3.5121165721
H             0.8947647665        3.6326334175        2.5577080920
O             7.9092632988        5.1778321931        2.9804596106
H             7.9302203121        4.3950128325        3.6368496793
H             8.8445883878        5.3735990087        2.7666863787
O             7.1490476346        8.7015176934        2.0872951493
H             6.4133313141        8.0992006975        1.8057962581
H             7.7311321482        8.1729782877        2.7000725866
O             8.8055944873        3.1706413860       -0.2451127771
H             9.2373131961        2.7162372312        0.5507813297
H             8.6295510165        2.4460364857       -0.8810804534
O             2.5592716829        1.3286407281        0.7432118330
H             2.5412687518        0.3662464963        0.9939262990
H             3.4229018113        1.6772690611        1.1203559007
O             4.9936600236        6.5071759499        4.1437667444
H             5.7593944236        6.9620921042        4.6125426587
H             5.1643747269        6.5239666366        3.1453767779
O             6.8395758039        7.6390960325        5.7557292322
H             7.6762233432        7.8349042076        5.2022212623
H             7.0988933644        6.9169659002        6.3696269439
O             2.3942019158        6.0614341205        1.0562854717
H             2.1185218796        5.4969362304        0.2757162330
H             2.1177239727        5.4757845906        1.8227538378
O             2.7376474776        7.7771223128        4.3639691553
H             3.6552717987        7.3168270615        4.3577941433
H             2.8137358276        8.6118213885        4.9011140507
O             2.9163861007        0.3430264818        5.4305296191
H             2.5873078132        0.5482589493        6.3309399761
H             3.9626433894        0.1976042254        5.5624886392
O             7.7248885700        3.1620677622        4.6776666908
H             6.7925154373        3.4621040474        4.9901877856
H             8.3127472053        3.4953711732        5.4107703366
O             1.5138595803        4.1504187250       -0.7692719357
H             0.5501292955        3.9274579632       -0.6701728919
H             1.9728339651        3.2957547931       -0.6069899492
O             2.1162855212        8.7095114416        1.8877275279
H             1.9942869086        7.8543420621        1.3850084014
H             2.5057005675        8.3720013553        2.7552817061
O             6.4635213544        4.2825834211        0.9852939811
H             7.0513848386        4.6013408537        1.7574491889
H             7.1414236974        3.9234500246        0.3526313951
O             8.8515678485        8.0143668802        4.1341927892
H             9.4157567545        8.8380795596        4.0460353697
H             9.5047025774        7.2584656056        4.2995603941
O            -0.9224592273        4.7957898006        6.5241277695
H            -0.2434039652        5.3115008347        5.9911151755
H            -0.4534280108        4.5188685696        7.3368748371
O             3.3181880062        2.8140713797        4.1322913626
H             3.8734353677        2.6445837079        3.2985477994
H             3.1103803235        1.9226874717        4.5274334387
O             5.4065815583       10.0542494512        5.7082343386
H             5.8545602695       10.3758871891        4.8592107719
H             5.8272885670        9.1717788333        5.8974369810
\end{verbatim}

\newpage

\section*{Surface \#7}
Tetrahedral cell $\alpha=\beta=\gamma=90^o$
\begin{verbatim}
    CELL_PARAMETERS (angstrom)
 10.0   0.0   0.0
  0.0   10.0   0.0
  0.0   0.0   20.0

ATOMIC_POSITIONS {angstrom}
O        0.492200879   2.925807549   4.931400324
H        0.405325271   2.672078325   5.862692365
H        0.944460553   2.148449717   4.461179092
O        7.537140074   6.420530242   4.254459592
H        8.046580639   5.621801977   3.900026709
H        6.749217320   6.039485046   4.725708414
O        9.004664597   6.176631291   0.343115708
H        8.464122023   5.353049025   0.308244048
H        9.859219219   5.838800709   0.697999148
O        3.202413788   7.001449900   4.991195465
H        4.004236069   6.511462422   5.313770722
H        3.458651540   7.428390574   4.114825409
O        0.478987467   7.399130986   5.239625298
H       -0.024828061   6.573606864   5.140568619
H        1.426858250   7.153886546   5.370284829
O        7.057499442   7.480790226   1.792353075
H        7.843085561   7.134362285   1.289939841
H        7.209290239   7.172752970   2.730495054
O        6.239219138   0.561762350   4.650605818
H        6.865625809  -0.061328011   5.103776900
H        5.369213324   0.481660090   5.117751867
O        4.547812034   3.448476768   4.042802015
H        3.644131053   3.772060711   3.806877450
H        4.950430834   3.062096437   3.222467128
O        5.579723375   9.862228693   2.250093328
H        6.235928961   9.195975072   1.934629082
H        5.887185944  10.101772900   3.203753671
O        3.541751057   8.065333719   2.577067459
H        4.209350404   8.807785283   2.471122500
H        3.907436126   7.312204108   2.024259430
O        2.142589058   4.732432908   3.735367019
H        1.523309816   4.221088315   4.319146017
H        2.432076095   5.547108330   4.220265077
O        8.529252242   1.374266797  -0.545721346
H        8.942634163   0.586586471  -0.104882866
H        7.692639476   1.049574533  -0.911944352
O        1.776208669   1.082282926   3.587979949
H        1.894788102   1.407625389   2.658282645
H        1.341085789   0.194245329   3.429735519
O        3.447699642   3.456515383  -0.029164836
H        2.748509367   4.047154811   0.384732914
H        3.863747276   3.990476658  -0.723808784
O        5.443444059   2.497389073   1.622637933
H        4.702215542   2.721561662   1.006687615
H        5.497928365   1.501468191   1.715088729
O        7.889900432   8.699091819   5.774101378
H        7.815499733   7.882429824   5.220697356
H        8.846218511   8.769323507   5.933764691
O        5.362906431   5.375890507   5.519710852
H        5.562497207   5.041498363   6.408210418
H        5.122420138   4.556882138   4.943957036
O        3.638775103   0.191218723   5.568958099
H        3.449506748  -0.764897029   5.568499332
H        3.094752727   0.563211939   4.835987540
O        7.836725039   3.598184050   0.879157042
H        8.205624768   2.821821371   0.376759455
H        6.933714186   3.269253049   1.137961265
O        8.876047381   4.409975810   3.252206739
H        8.507868564   3.999541580   2.422774526
H        9.208971728   3.696542942   3.837970816
O        1.334070017   4.631009055   1.148300003
H        0.862294989   3.774774692   1.123432866
H        1.547171968   4.762225538   2.114980905
O        1.573483558   1.622383408   0.894486300
H        1.177634787   0.815688545   0.501435160
H        2.261476114   1.946849085   0.277917772
O       -0.076076686   9.244069058   0.460295316
H       -0.262977774   8.414936567  -0.015150182
H        0.126464009   8.961343368   1.395727067
O        0.915118608   8.654129308   2.864615360
H        0.626015950   8.135489834   3.662909704
H        1.852397118   8.370717061   2.666726523
O        4.801179805   6.113213171   1.220358757
H        5.697755694   6.543873623   1.363137589
H        4.841092072   5.272910977   1.706898806
\end{verbatim}

\newpage

\section*{Surface \#8}
Tetrahedral cell $\alpha=\beta=\gamma=90^o$
\begin{verbatim}
    CELL_PARAMETERS (angstrom)
 10.0   0.0   0.0
  0.0   10.0   0.0
  0.0   0.0   20.0

ATOMIC_POSITIONS {angstrom}
O        5.028978576   4.177949891   1.058381269
H        5.574957392   4.936821133   1.401234756
H        5.639083942   3.637087586   0.527418034
O       10.291254128   7.158431835   3.863308576
H        9.860180508   6.551495374   3.200155641
H       10.931488227   6.597934570   4.366894975
O        1.348968607   2.801616923   4.776013971
H        2.249386510   2.401861518   5.030097596
H        1.246502720   2.563100217   3.799304136
O       -0.523456016   1.114952613   5.672965541
H        0.196976947   1.764947157   5.373321312
H       -0.692358002   1.337331818   6.602779628
O        3.694098203   1.733578169   5.276909071
H        3.995682690   1.718559784   4.311873926
H        4.348850400   2.352010465   5.683198402
O        0.655418181   7.247909460   0.338203613
H        0.462057845   7.495177008  -0.579724488
H        1.603531712   6.900538798   0.335183178
O        5.742823166   3.686245128   5.611031270
H        5.366903626   4.369233573   5.019303082
H        6.330195245   3.134117484   5.030211629
O        4.111524032   8.333045960   2.577643392
H        3.263501097   8.841921582   2.495754229
H        4.011443113   7.757631860   3.368428766
O        9.008386325   5.710927766   1.929355930
H        8.080496761   6.044214700   1.981889488
H        9.481111878   6.208239355   1.209949386
O        6.625944491  -0.400984311   2.948660892
H        7.143716244  -0.943417516   3.593476986
H        5.694947826  -0.730114506   2.943428180
O        1.443067621   9.228295639   2.273943753
H        1.088815835   8.650798040   2.994145431
H        1.151981247   8.751621775   1.461437204
O        4.541525837   1.884338819   2.803626169
H        4.681643297   2.772579539   2.418205763
H        4.332377027   1.307905877   2.007361453
O        4.142898602   0.595142308   0.488983647
H        3.590881728   1.328385305   0.115029568
H        5.068996378   0.754582159   0.191956834
O        8.520346238   2.949186199   1.752044138
H        8.072531447   2.748826062   2.618574160
H        8.711000528   3.920690148   1.752473456
O        7.354051188   2.060030885   4.065499994
H        8.076141581   1.698546512   4.635395741
H        6.877518018   1.270396025   3.693313990
O        2.270375202   9.214737091   5.450900183
H        2.887309003   9.975832711   5.522909571
H        1.396929525   9.616078866   5.596906870
O        6.386078899   6.824179170   6.222585212
H        6.762214933   5.936582371   6.358906036
H        7.110377074   7.395058639   5.858546331
O        6.279557404   6.302768204   2.194548696
H        6.019889119   7.175421000   1.844890995
H        5.732067772   6.210705956   3.023027942
O        4.626598183   6.171014637   4.354531105
H        3.722598749   5.885019050   4.645499109
H        5.149823355   6.506318761   5.145488695
O        8.251088118  -1.520740444   4.943389102
H        9.038599967  -2.041380333   4.597113023
H        8.637091459  -0.688612285   5.286873963
O        0.996779834   1.975621588   2.252828949
H        0.059248358   2.217760936   2.002043162
H        1.077718758   0.985318311   2.251476982
O        3.213964289   6.628121386   0.478805619
H        3.657516582   5.761744266   0.522951134
H        3.622303345   7.198974421   1.174776845
O        2.555127498   2.841975986   0.099661056
H        3.216985660   3.451100422   0.474951495
H        1.984152258   2.589305663   0.868698054
O        6.956719239   0.907859092   0.415575023
H        7.506875305   1.686311522   0.646903561
H        6.952386466   0.383832126   1.247904108
O        2.062728836   5.390133486   5.142552007
H        2.014685028   5.539841459   6.100917698
H        1.757359576   4.438773783   4.994863748
\end{verbatim}

\newpage

\section*{Surface \#9}
Tetrahedral cell $\alpha=\beta=\gamma=90^o$
\begin{verbatim}
    CELL_PARAMETERS (angstrom)
 10.0   0.0   0.0
  0.0   10.0   0.0
  0.0   0.0   20.0

ATOMIC_POSITIONS {angstrom}
O        2.312679622   5.464990674   3.359398197
H        2.225679579   6.335356554   3.812213595
H        1.652480289   4.830838437   3.763207022
O        5.517028555   7.670836513   4.119779746
H        5.501593564   6.733986034   3.760781344
H        5.911265882   7.627848637   5.014365598
O        7.059261516   1.067677534   1.447498063
H        7.835581700   1.417456594   0.936912427
H        7.422318071   0.632700569   2.256672443
O        9.125536533   4.752817693  -0.054115796
H       10.022001433   5.090206449   0.147197066
H        8.489384172   5.318029508   0.464348080
O        1.822174250   5.649303995   0.663741022
H        2.031228796   5.611787212   1.641286177
H        2.505044025   6.210779334   0.259652302
O        8.351849362   5.385746827   4.127266083
H        7.676640639   4.909332539   4.674258943
H        9.100647690   4.732279098   4.027616207
O        1.721488166   9.453542060   2.331082630
H        2.433916037   9.482639430   1.605563202
H        2.117211832   8.986069809   3.103097907
O        7.668413074   6.526324567   1.524637176
H        7.731796585   6.105203978   2.410948737
H        8.463105051   7.136590993   1.475755770
O       -0.320060918   1.998934953   6.193272013
H       -1.268047722   1.706169646   6.093105144
H       -0.250708908   2.325963253   7.102983872
O        3.071656443   2.783759800   0.540061633
H        2.742422579   3.647166194   0.225072389
H        2.375296509   2.502552457   1.195961541
O        0.995715885   1.988389510   2.127178170
H        1.214616930   1.029661771   2.330223939
H        0.308712757   1.976998999   1.403326619
O        9.160796287   2.186955785   0.159811774
H        9.395561919   1.821280980  -0.707367311
H        9.082471868   3.204742346   0.050586007
O        7.062858329   1.177163499   5.800988626
H        6.642164973   2.056406619   5.631346417
H        7.186379875   0.712967698   4.920889862
O        5.054079353   5.085542425   3.404681701
H        4.062797274   5.101090267   3.393276621
H        5.322589077   4.471344001   2.628351283
O        0.454776657   3.662031610   4.106687902
H        0.259945399   3.115594630   4.911800361
H        0.603475982   2.992404015   3.367505082
O        4.718996096   6.066297182   6.844109449
H        5.036505008   6.988531332   6.913311451
H        4.044920679   6.099215135   6.144277345
O       -0.653149145   7.819348293   4.767182164
H       -1.018411609   6.902499480   4.586988967
H       -0.758254409   7.946134722   5.723004110
O        5.580447559   3.428018382   1.485288611
H        4.696128651   3.137099998   1.141539582
H        6.109679743   2.596659285   1.586073906
O        6.145840946  -1.296420490   6.787619299
H        6.674663747  -1.385290859   7.595375871
H        6.317168779  -0.363100811   6.488910293
O        3.415672349  -0.403045505   0.350333861
H        3.553741207   0.530541653   0.108432453
H        4.323494450  -0.805971442   0.486964238
O        7.533730287   9.577796543   3.693787261
H        6.784846336   8.940814818   3.592971772
H        8.290468163   9.000901720   3.995625046
O        5.864982030   8.630868321   0.706311282
H        6.406338485   9.434451323   0.890572353
H        6.377718688   7.848270022   1.011616026
O        2.859560359   7.987019399   4.424968564
H        3.861773338   7.990353207   4.347398194
H        2.626119774   8.448529781   5.245268924
O       -0.118713707   7.914062754   1.059352467
H        0.491587552   7.204923200   0.778602481
H        0.459394337   8.508660862   1.607390219
O        6.311924835   3.939291274   5.451049955
H        5.943789999   4.462487703   6.192798988
H        5.745618777   4.249035738   4.676218668
\end{verbatim}

\newpage

\section*{Surface \#10}
Tetrahedral cell $\alpha=\beta=\gamma=90^o$
\begin{verbatim}
    CELL_PARAMETERS (angstrom)
 10.0   0.0   0.0
  0.0   10.0   0.0
  0.0   0.0   20.0

ATOMIC_POSITIONS {angstrom}
O        9.811876652   8.118569040  -1.112946991
H        9.188412710   8.810744691  -0.846412403
H       10.711704839   8.538450965  -1.128791057
O        4.776948758   2.788129414   0.489846646
H        5.583764631   2.805745269   1.046972594
H        4.020721951   2.823129144   1.123514748
O        4.457199579   0.270922504  -0.566279850
H        4.529190958   1.229362495  -0.307548178
H        4.756245378  -0.214408381   0.241796231
O        7.080648916   2.995649786   2.383655012
H        6.865675822   3.754765250   2.987216780
H        8.063635907   3.025191245   2.254684353
O       10.478406673   8.181942099   3.297021999
H       10.763472966   7.673111137   4.106877502
H       11.233457434   8.128468079   2.643798206
O       -0.576026725   5.684111946  -0.034480806
H       -0.362740674   6.549162285  -0.488122131
H        0.178053346   5.090718159  -0.171559906
O        6.600086964   4.713528848   4.519043852
H        7.562289920   4.663507280   4.775798027
H        6.054691709   4.257679170   5.209158971
O        3.461314073   5.903032595   5.316090350
H        3.591043388   5.809370043   4.335025218
H        3.802298751   5.065210090   5.701555052
O        8.676104423   6.430926738   2.370714295
H        9.050586616   6.092907077   1.496417007
H        9.276385246   7.190299847   2.634304509
O        5.016758750  -0.453848881   4.183109075
H        5.781608507   0.181687654   4.275938998
H        5.370575313  -1.297716978   4.578025462
O        0.998584994   6.655017836   5.512532948
H        0.860705562   7.129123786   6.347835008
H        1.966958297   6.321745312   5.524957078
O        4.789901794   3.484237663   6.114266290
H        4.706371481   3.265429574   7.054765409
H        4.287908479   2.765918233   5.588398252
O       -0.749313400   4.801131702   4.425391051
H       -0.117068053   5.288946205   5.007588297
H       -0.916973387   5.425481190   3.660561002
O        7.024125840   1.290651197   4.419954823
H        6.871997113   1.858581244   5.194550062
H        7.001235491   1.911697046   3.620975358
O        2.466489043   7.773170285   1.489290952
H        2.291959945   7.935238122   0.530628458
H        3.353927990   8.238587173   1.619645421
O        6.122377119   6.266113754   1.514202059
H        7.037038504   6.349733307   1.911613066
H        6.249257569   5.782252917   0.680885530
O        2.434283406   8.772007950  -1.144439704
H        3.146623184   9.484868887  -0.940669435
H        2.777406395   8.299891863  -1.919450978
O       -0.036794619   2.863503403   2.656095419
H       -0.072126863   1.983169896   3.132686955
H       -0.147264029   3.520800159   3.396844135
O        6.553202511   7.444242471   4.869891849
H        6.415664512   6.466728195   4.758705846
H        7.365671320   7.625677399   4.370780243
O       -0.301531285   0.643573622   4.158280030
H       -0.031370010  -0.255648852   3.834188620
H       -1.275649966   0.655478045   4.302105021
O        1.048024831   2.133672981   6.252399878
H        0.493978727   1.570937341   5.673206400
H        1.924057401   2.032553752   5.828030620
O        3.660541309   5.583064689   2.615323061
H        4.592200987   5.691386080   2.299082400
H        3.145077311   6.279925615   2.126009764
O        4.907732281   8.759464656   1.669815403
H        5.441522141   7.932328871   1.584664181
H        5.008521134   9.096116768   2.615636606
O        2.731179809   3.026016048   2.421612267
H        1.735490453   2.983995993   2.393690317
H        2.975141481   3.985666152   2.517448708
O        3.467884732   1.776969522   4.695341267
H        3.211233837   2.178236029   3.813281442
H        3.854137889   0.881376397   4.536285907
\end{verbatim}